\documentclass{article}

\usepackage[main, final]{iaseai26}

\usepackage[utf8]{inputenc} 
\usepackage[T1]{fontenc}    
\usepackage[svgnames,x11names,table]{xcolor} 

\usepackage[colorlinks=true, urlcolor=DarkBlue, linkcolor=black, citecolor=black]{hyperref}

\usepackage{url}            
\usepackage{booktabs}       
\usepackage{amsfonts}       
\usepackage{nicefrac}       
\usepackage{microtype}      
\usepackage{xcolor}         
\usepackage{graphicx}
\usepackage{csquotes}
\usepackage{todonotes}      

\usepackage{xcolor}
\usepackage{enumitem}
\setlist[itemize]{leftmargin=0.5cm}
\usepackage{titlesec}
\usepackage{parskip}
\usepackage{wrapfig}

\title{An International Agreement to Prevent the Premature Creation of Artificial Superintelligence}

\author{%
Aaron Scher\thanks{Equal contribution.} \quad
David Abecassis\footnotemark[1] \quad
Peter Barnett\footnotemark[1] \quad
Brian Abeyta\footnotemark[1] \\
Machine Intelligence Research Institute \\
Technical Governance Team\\
\texttt{\{aaron.scher,david,peter,brian\}@intelligence.org}
}

\begin{document}

\maketitle

\begin{abstract}
Many experts argue that premature development of artificial superintelligence (ASI) poses catastrophic risks, including the risk of human extinction from misaligned ASI, geopolitical instability, and misuse by malicious actors. This report proposes an international agreement to prevent the premature development of ASI until AI development can proceed without these risks. The agreement halts dangerous AI capabilities advancement while preserving access to current, safe AI applications.

The proposed framework centers on a coalition led by the United States and China that would restrict the scale of AI training and dangerous AI research. Due to the lack of trust between parties, verification is a key part of the agreement. Limits on the scale of AI training are operationalized by FLOP thresholds and verified through the tracking of AI chips and verification of chip use. Dangerous AI research—that which advances toward artificial superintelligence or endangers the agreement’s verifiability—is stopped via legal prohibitions and multifaceted verification.

We believe the proposal would be technically sufficient to forestall the development of ASI if implemented today, but advancements in AI capabilities or development methods could hurt its efficacy. Additionally, there does not yet exist the political will to put such an agreement in place. Despite these challenges, we hope this agreement can provide direction for AI governance research and policy.

\end{abstract}
\section{Introduction}
The 2025 book \textit{If Anyone Builds It, Everyone Dies} argues that the development of artificial superintelligence (ASI, AI substantially smarter than humans at all cognitive tasks) “using anything remotely like current techniques, based on anything remotely like the present understanding of AI” would lead to human extinction~\citep{yudkowsky2025ifanyone}. If that’s the case, what should we do about it? Experts differ in their assessment of the situation. Many think that there is around a 10\% chance of future AI advances causing human extinction (for aggregations of expert opinion see \cite{grace2025thousands, pdoom2025wikipedia, controlai_quotes_2025}). Even if there were “only” a 10\% chance of human extinction, how should governments respond?

In late 2025, tens of thousands of researchers, public figures, and members of the public signed a statement calling for a prohibition on the development of superintelligence until there is ``broad scientific consensus that it will be done safely and controllably, and strong public buy-in''~\citep{superintelligence_statement_2025}. This report describes what such a prohibition could look like in practice: an international agreement which, if implemented, would postpone the development of ASI until proper mitigation efforts are developed and implemented (i.e., potentially decades). Practically speaking, this agreement is not going to be implemented tomorrow, but we hope it can serve as an end goal for AI governance, and in Appendix~\ref{app:stages} we lay out a staged implementation that iteratively builds up to the final agreement.

Our proposed agreement is not without tradeoffs and risks: it forgoes beneficial AI capabilities, creates some pathways for authoritarian risk (although we think less risk than the current trajectory), cedes the U.S. technological lead in AI, risks leaking sensitive information due to verification, and would impose costs on industry and financial markets. Despite these downsides, we believe this agreement is worth pursuing because the alternative of racing towards ASI poses even greater risks.

Contributions:
\begin{itemize}[nosep]
    \item We provide a brief overview of AI risks, and an overview of the key details of the strategic situation informing our proposal.
    \item We describe the implementation of an international agreement to forestall the development of ASI until such a time as it can be developed safely.
    \item We explain the key assumptions and beliefs that differentiate our plan from others in the space.
    \item We provide the full text of an example agreement, detailed commentary, and connections to existing agreements in Appendix~\ref{app:agreement}. We discuss immediate steps different actors could take toward this agreement in Appendix~\ref{app:whatcanwedo}. We explain a staged implementation of the agreement that slowly builds from the current world to the agreement in Appendix~\ref{app:stages}. We detail various approaches that could help locate existing AI chips and bring them under monitoring in Appendix~\ref{app:locatingchips}.
\end{itemize}

\section{An overview of AI risks}
\label{sec:overview_of_ai_risks}
There are many large-scale risks from advanced AI (i.e. AI much more powerful than today’s AI technology) that might warrant slowing or stopping AI progress for a significant period (e.g., decades). Most of these risks stem from future AIs becoming increasingly capable tools: Rapid automation may destabilize economies \citep{drago2025intelligence,erdil2025gate}. Capabilities that undermine strategic deterrence and incentivize first strikes may trigger wars and conflicts \citep{PE-A3691-4,hendrycks_superintelligence_2025}. Malicious actors may misuse AI to engage in wide-scale cyberattacks or to develop novel biological weapons \citep{zelikow2024defense, crawford2024securing, esvelt2022delay}. AI used for mass surveillance, propaganda, and police-state oppression may enable extreme concentrations of power \citep{davidson2025aienabledcoupsho, bullock2025agi, drago2025intelligence}.

But while the above risks may lead to unprecedented catastrophes, humanity can probably survive them. A greater and more irreversible risk stems from AI systems that don’t remain passive tools, but instead become capable enough to permanently disempower humanity. This would plausibly lead to human extinction, either as an intentional act by the AIs (e.g., to prevent humans from making powerful competing AIs) or as a side effect of AIs optimizing for goals that don’t specifically designate human survival \citep{yudkowsky2022agi, ngo_alignment_2025}.

This paper focuses on an international agreement to forestall ASI development until it can be done safely. A halt like this is primarily motivated by concerns about misalignment and AI takeover, and we think it is the only acceptable mitigation to those risks. But a halt would also be an effective tool for addressing many other risks, such as geopolitical destabilization, mass AI-caused unemployment, and misuse of powerful AIs \citep{barnett2025ai}. Therefore, \textit{others may support a halt even if they are not concerned with misalignment}.

Our view is that AI developers are not on track to solve the alignment problem before the development of ASI. We will not fully defend this claim here—the difficulty of alignment is not this document's central focus. Instead, we provide some general intuitions for why AI alignment appears challenging enough that success is far from guaranteed. This is sufficient to motivate our policy proposal.

The deep learning paradigm underpinning modern AI development seems highly prone to producing agents that are not aligned with humanity’s interests. Under this paradigm, AI behavior is not hand-coded by human engineers, but instead emerges from automated training processes. Rather than producing lines of code that can be debugged when something goes wrong, these processes configure hundreds of billions to trillions of model weights that determine AI behavior. While it is sometimes possible to identify and activate features of these weights correlated with certain inputs or outputs, the deep machinery of these minds is effectively opaque to human observers~\citep{bereska2024mechanistic}. No one can say with any certainty why an AI made a particular choice or how it would act in situations that differ from its training.

Consequently, no one can manually adjust an AI’s weights to give it a particular goal or set of preferences. And yet, as seen in the growing ability of AIs to complete long-duration tasks, AIs are becoming increasingly capable of pursuing goals, e.g., in the setting of agentic software engineering \citep{kwa2025measuring}. Various well-documented examples of misbehavior by these same models (both in the lab and in the wild~\citep{baker2025monitoring, bondarenko2025demonstrating, lynch2025agentic, openai2025sycophancy, metr2025reward}) demonstrate that today’s training methods do not produce robustly aligned agents that could be trusted not to disempower humanity if they had the capacity.

Many AI labs have a stated goal of building AIs that can surpass humans at all cognitive tasks. It is hard to predict \textit{when} AI labs might succeed, but many predict ASI could be developed soon. The field has seen enormous progress in just five years—from AIs that couldn’t hold a conversation to “reasoning models” that solve hard exam questions \citep{EpochLLMBenchmarkingHub2024} and that some major tech companies claim write a substantial fraction of their code~\citep{chandra2024ai, novet2025satya}. Another few years of such progress might well push capabilities well beyond human level in all or nearly all domains.

One accelerating factor complicating all forecasts is the possibility of AIs automating the process of AI research and development itself \citep{davidson2025threetypesofinte,eth2025willairdautomati,wijk_re-bench_2024}. The distance to this threshold is unclear, but some independent experts give median prediction dates ranging from 2028 to 2033 \citep{lifland2025timelines}. If or when this threshold is crossed, we may see a feedback loop where AI R\&D is conducted by AIs that rapidly (in months or a few years) produce substantially superhuman AI~\citep{openai_openai_2025, anthropic_rsp_2025, google_frontier_safety_2025}.

We note that artificial general intelligence (AGI)---systems that perform at human-level across all cognitive domains---would likely emerge before ASI. While AGI itself poses significant risks, our primary focus is preventing ASI, as misaligned superintelligence poses an irreversible existential threat. However, AGI capable of automating AI research could accelerate the path to ASI and trigger rapid, uncontrollable capability gains. This agreement aims to halt AI capabilities advancement before the creation of AGI systems that could automate AI research.

On the current trajectory, the core challenge of AI alignment is unlikely to be adequately addressed before the first ASIs are built. Evidence for this can be seen in the failure to solve far easier related problems. Today’s frontier models are not robust to jailbreaks: It is relatively easy to craft inputs that lead to outputs that AI developers attempt to avoid \citep{andriushchenko2024agentharm,andriushchenko2024jailbreaking,zou2023universal,anil_many-shot_2024,wang_jailbreak_2024}. Current models, especially reasoning models, display a propensity to reward hack (exploiting unintended solutions) and lie to their users, against the best efforts of their developers \citep{bondarenko2025demonstrating,wijk_re-bench_2024,chowdhury2025investigating,anthropic2025claude37,baker2025monitoring}. Far tougher alignment challenges are likely to appear only at advanced capability levels where humanity would be unable to recover from failures \citep{yudkowsky2022agi}.

There is a broad consensus in the field that catastrophic outcomes from misaligned AI pose a significant threat that should be addressed \citep{bengio2024managing}. Hundreds of luminaries signed on to a 2023 statement that “Mitigating the risk of extinction from AI should be a global priority alongside other societal-scale risks such as pandemics and nuclear war.” \citep{cais_statement_2023} That same year, Yoshua Bengio, a Turing Award recipient and the most cited living scientist, said he thinks there is a “twenty percent probability that it turns out catastrophic” \citep{lavoipierre2023ai}. In a 2023 survey, 38\% of survey respondents at top AI conferences said there was at least a 10\% chance that AI’s outcome will be "extremely bad (e.g., human extinction)", conditional on the development of AI that can outperform humans on all tasks~\citep{grace2025thousands}.

This understanding extends to AI company leaders themselves: Anthropic CEO Dario Amodei has predicted a 10-25\% chance that “something goes really quite catastrophically wrong on the scale of human civilization” \citep{bartlett2023anthropic, amodei_axios_2025}. OpenAI’s Sam Altman has called superintelligence “probably the greatest threat to the continued existence of humanity” \citep{altman2015machine}, and xAI’s Elon Musk has said, “AI is a fundamental risk to the existence of human civilization” ~\citep{london2017musk}. Musk has indicated that his participation in this ASI race stems from the belief that it would happen even without him \citep{musk2025interview}. Those racing toward superintelligence, to the extent they are concerned with catastrophic risks, are stuck in a collective action problem \citep{armstrong2016racing}. Halting the race and avoiding extinction would require \textit{global coordination encompassing all relevant companies and governments}.

\section{The strategic situation}

A 10\% chance of extinction is unacceptable. To mitigate these risks, various approaches have been proposed: Have governments tightly regulate and license AI development~\citep{anderljung2023frontier, smith2024licensing}, empower defensive actors early \citep{buterin_my_2023,nielsen_notes_2024,bernardi_societal_2025}, implement a joint global AI development project to avoid racing \citep{brundage_my_2024,hausenloy_multinational_2023}, race ahead while slowing opponents \citep{aschenbrenner_situational_2024,amodei_machines_2024}, or prevent the creation of ASI before we are ready \citep{barnett2025ai, ramiah2025toward, aguirre2025keep}.

We think the only acceptable solution is to prevent the creation of ASI until catastrophic risks are mitigated. The state of AI alignment research is too nascent, and the problem too difficult. Nobody knows how to align advanced AIs to human values, including regulators or an international standards body (and we address some of these alternative plans in more depth in Section~\ref{sec:whythisplan}). Therefore, we think the right approach to this problem is to forestall the development of ASI for as long as is necessary to solve critical alignment challenges. We note that bans are not usually the best way to address dangerous technologies, but ASI is uniquely dangerous. An agreement establishing such a halt would need to be globally enforced, verifiable, and ready to implement once the political will to do so exists.

For a halt to be effective, it must extend to all countries. If only cautious actors stop, other parties will continue the march toward superintelligence. The United States (U.S.) pausing on its own might buy around six months until the nearest company in the People’s Republic of China (PRC) catches up, and a U.S.–PRC agreement without global enforcement would similarly fall to the next closest actor, likely in a couple of years \citep{artificialanalysis2025countries} (though it is unclear how long it would take to get from the current frontier to ASI). The U.S. and PRC are the preeminent superpowers and the centers of most frontier AI development; therefore, they are essential stakeholders in such a deal, and their strategic interests must be accommodated. Due to their influence, they are likely leaders in future AI agreements, whether or not other countries participate as well.

A coalition including the U.S. and PRC would likely be able to attract nearly all other states, even if some are reluctant, given these two countries’ geopolitical influence and their current positions as AI leaders. We expect other countries to sign on to such an agreement once they understand AI risks better and conclude that international coordination is necessary to avoid these risks. Participation could also be encouraged with benefits such as security and access to AI infrastructure for allowed use. Where stronger incentives prove necessary, standard tools such as economic sanctions may apply. We believe a very ASI-concerned coalition headed by the U.S. and PRC would be sufficient to integrate other potential AI developers into the agreement or to sufficiently suppress other AI development so that it does not pose a threat.

All parties to the agreement will desire assurance that other parties are not engaged in secret AI capabilities projects. Providing this assurance is a key aim of the agreement. Given a low-trust environment, parties must have the means to verify compliance to their satisfaction. It gives special consideration to the independent verification efforts of the U.S. and the PRC.

There remain significant political obstacles to such an agreement occurring today. Nevertheless, to accept even a 10\% chance of extinction is wholly inconsistent with how we manage other risks. We routinely adopt stringent mitigations for risks that are hundreds of times less likely and that would result in millions of times fewer fatalities, such as in the nuclear industry, and in many other fields \citep{koessler2024risk,nrc2017approach, ehrhart2020evaluation}.

In addition, the rate of progress is volatile and uncertain. Timelines to ASI may be short, and some possible developments in AI would preclude a robust monitoring and verification regime. Rapid feedback loops, hardware proliferation, and the loosening of supply chain bottlenecks all become more likely the longer an arrangement is delayed.

When world leaders are mobilized to address the situation, it is essential to have prepared measures for negotiation and rapid adoption by key players and the world at large.

\section{The Agreement}
\begin{figure}[t]
    \centering
    \includegraphics[width=0.6\linewidth]{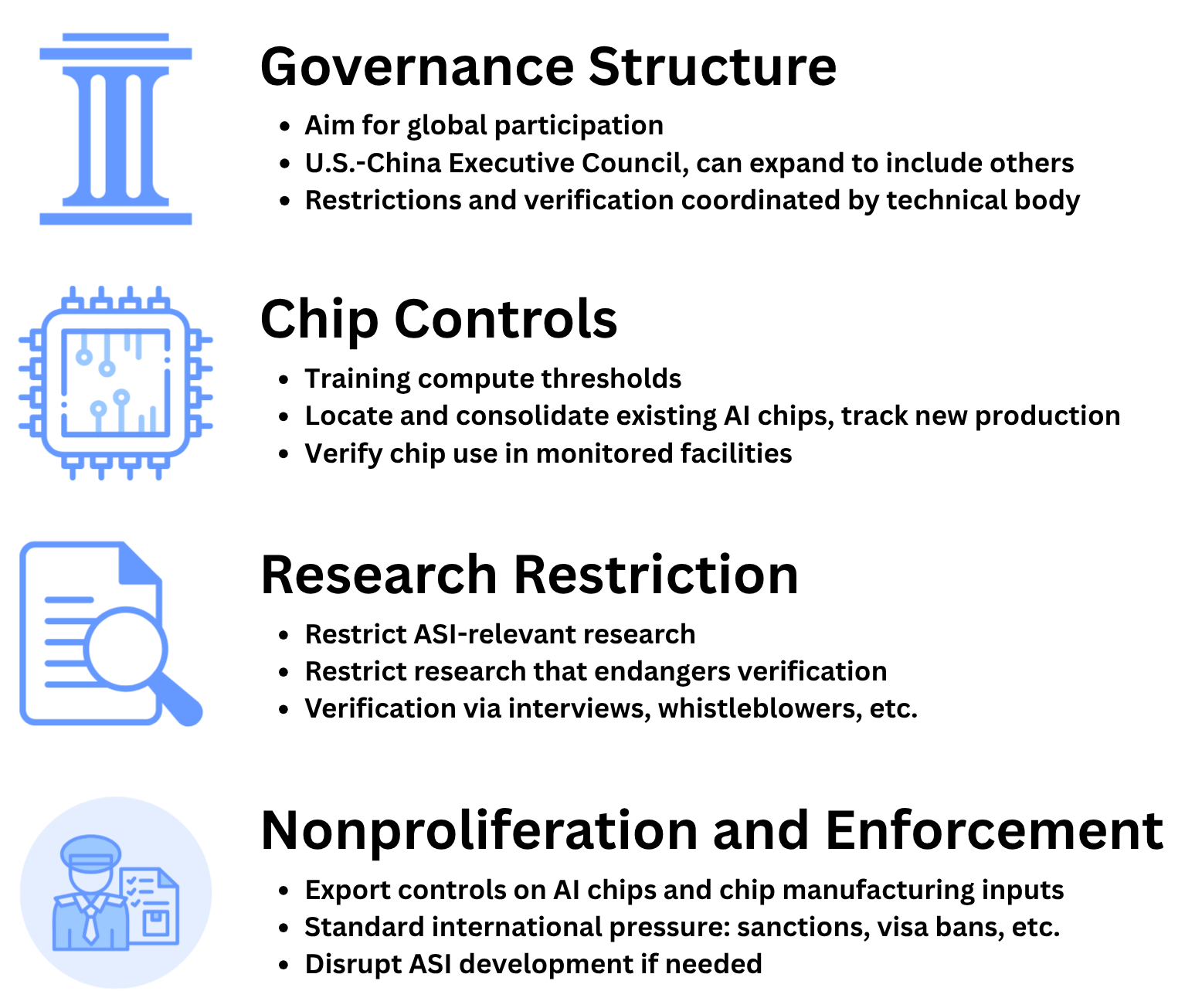}
    \caption{An overview of the agreement’s main components.}
    \label{fig:agreement_overview}
\end{figure}


\textit{A full draft agreement, with discussion of precedent and other considerations, is presented in Appendix~\ref{app:agreement}. For concision, this section describes how an implementation of that agreement might look in practice. Figure~\ref{fig:agreement_overview} provides an overview.}

Motivated by the risks above, the U.S. and PRC lead a coalition of states to prevent the advancement of AI capabilities toward ASI. An agreement establishes how the U.S. and PRC coordinate to gain assurance that neither they nor anyone else is advancing AI capabilities. It empowers and accommodates states’ information gathering efforts and, in particular, supports independent efforts by the U.S. and the PRC to verify compliance.

Decision-making power over the agreement reflects the underlying geopolitical realities: Today and for the foreseeable future, the U.S. and PRC are the primary actors whose participation is key to any AI agreement. They are the initial members of an Executive Council which governs the agreement. This body can be expanded as appropriate and necessary, and we anticipate that other states with significant power or specific leverage over AI development may be included. That said, the responsiveness and flexibility of the agreement is maximized by limiting the number of parties required to modify it. The coalition nevertheless expands to include as many countries as possible over time.

AI development in the current deep learning paradigm is driven both by the large-scale employment of specialized computer chips and by improvements in AI algorithms \citep{epoch2024trainingcomputeoffrontieraimodelsgrowsby45xperyear,kaplan_scaling_2020, ho_algorithmic_2024}. Therefore, the coalition stops the advancement of AI capabilities toward superintelligence by restricting the scale of AI training (operationalized as total training FLOP) and by restricting research into AI algorithms and other areas that would undermine the governance regime. These research restrictions are narrowly targeted at research that advances towards ASI or would undermine the agreement.

AI training runs above the Strict Threshold (i.e., $10^{24}$ FLOP) are prohibited. Training runs below this threshold but above the Monitored Threshold (i.e., $10^{22}$ FLOP) must be approved and monitored by coalition authorities. Training runs below the Monitored Threshold require no approval or monitoring.

\begin{figure}
    \centering
    \includegraphics[width=1\linewidth]{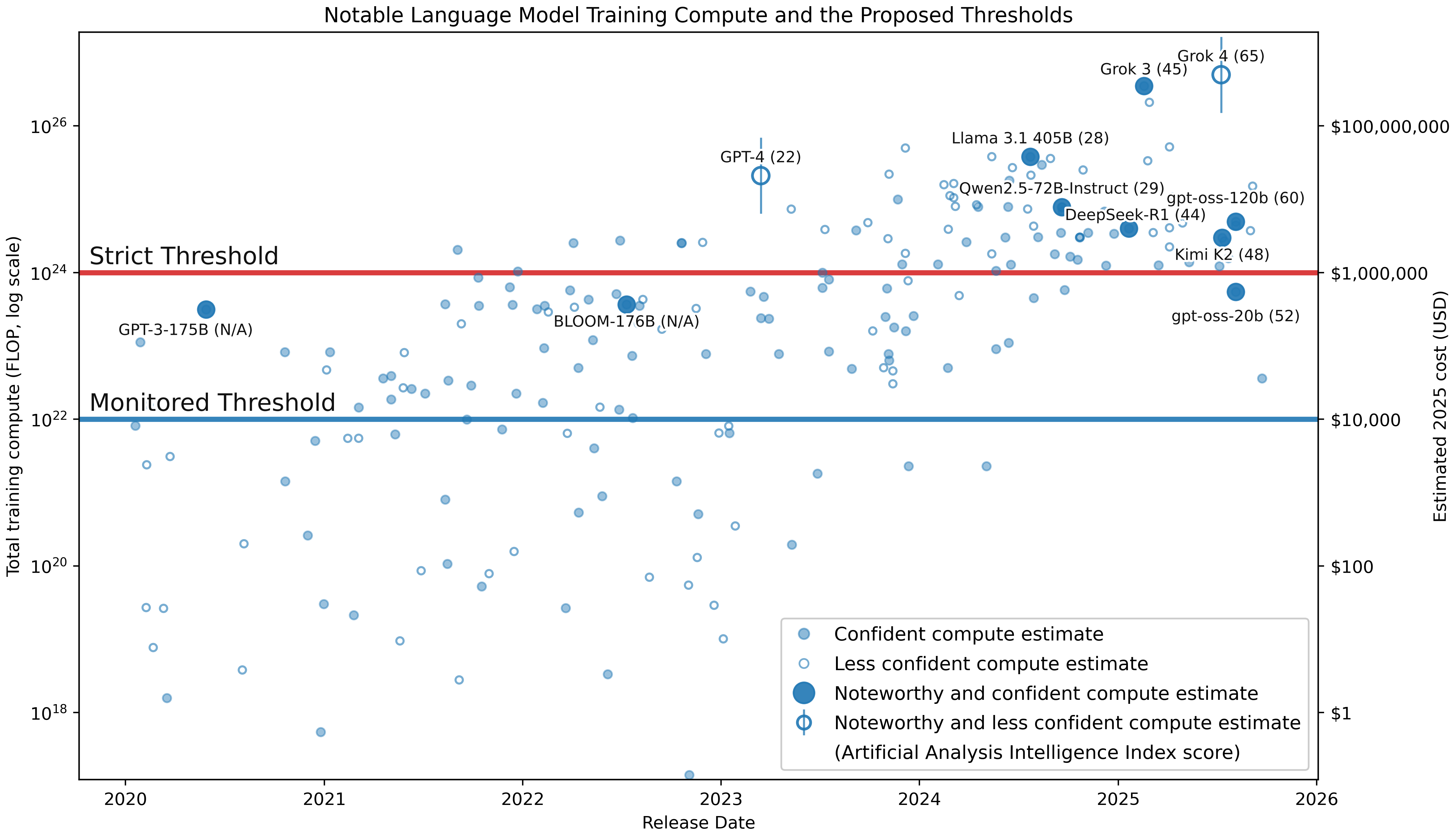}
    \caption{The training compute used to train various notable AI models in the last few years, along with the two thresholds in this agreement. In the agreement, new training above the Strict Threshold ($10^{24}$ FLOP) would be prohibited and new training between the Monitored Threshold ($10^{22}$ FLOP) and Strict Threshold ($10^{24}$ FLOP) would be monitored. Due to a lack of public data, only some models have confident estimates for the training compute used. The number in parentheses is the model’s score on the Artificial Analysis Intelligence Index. Data is based on \cite{EpochAIModels2025} and \cite{artificialanalysis_models}. Cost estimates are based on October 2025 rental prices for B200 GPUs from CoreWeave, assuming 50\% utilization in FP8.}
    \label{fig:models_and_training_compute}
    \vspace{-0.3cm}
\end{figure}

There are numerous considerations in deciding where these thresholds should be set, and these thresholds can be modified during agreement negotiations or after the agreement is in effect. First, nobody knows what scale of training would reach particularly dangerous AI capability levels; given this uncertainty and the massive negative effect of setting thresholds too high, we suggest a conservative approach. Second, many near-frontier AI models of today are trained at a scale only slightly above the Strict Threshold; for example, DeepSeek-R1 ($\sim4 \times 10^{24}$ FLOP) and gpt-oss-120B ($\sim5 \times 10^{24}$ FLOP)~\citep{EpochAIModels2025}. See Figure~\ref{fig:models_and_training_compute} for a visualization of how these thresholds relate to the training scale of existing AI models. Due to improvements in AI algorithms and data, the capability of models trained at a given computational scale increases rapidly over time \citep{ho_algorithmic_2024}. Due to likely progress in algorithms and data between today and when this agreement would come into effect, AIs trained at the Strict Threshold will be more capable—potentially much more—than the models trained at that scale today. There are numerous other considerations in what these thresholds should be and whether FLOP thresholds are appropriate~\citep{erben_JRC143255, hooker2024limitations, heim_training_2024}; a full discussion is out of scope for this paper.

Verification of these restrictions is practical because AI chips are expensive and specialized, and thousands of them are needed for frontier AI development. This approach helps ensure that training runs that advance toward ASI are prevented while still allowing continued use of current AI technology.
To achieve this verification, the coalition \textbf{locates existing AI chips}, \textbf{tracks new AI chips being produced}, and \textbf{monitors AI chip use} in order to ensure chips are not being used for restricted activities \citep{scher_mechanisms_2024, baker2025verifying, harack2025verification}. Chips are located through coordinated efforts by intelligence agencies and domestic law-enforcement agencies from coalition countries. These efforts involve combining supply chain tracking, mandatory reporting, state intelligence gathering, open-source intelligence, power consumption monitoring, challenge inspections, whistleblowers, and more. Appendix~\ref{app:locatingchips} provides more detail on how this could be done. Tracking new AI chips relies on the concentrated chip supply chain and serves the goal of making sure chips are not smuggled outside of the monitoring regime. Finally, the chips themselves are monitored in order to verify that they are only being used for acceptable use cases.

AI chip monitoring starts with discerning whether the chips are being used for inference on existing AIs or training of new AIs. The coalition works to develop tamper-resistant on-chip mechanisms for such purposes, as finer-grained control permits greater use of AI chips for safe applications, with fewer inspections. Initially inspectors are given ongoing physical access to chips, as is likely needed for robust chip-use verification \citep{aarne2024secure,kulp2024hardwareenabled}. To ensure chip use verification is applied, the coalition prohibits large concentrations of chips (i.e., greater than 16 H100-equivalents; 16 H100s cost approximately \$500,000 USD in 2025) outside of monitored facilities.

Despite prohibitions on large-scale training, AIs could continue getting more capable via improvements to the algorithms and software used to train them. Therefore, the coalition adopts appropriate \textbf{restrictions on research that contributes to frontier AI development} or research that could endanger the verification methods in the agreement. These restrictions would cover certain machine learning-related research and may eventually expand to include other AI paradigms if those paradigms seem likely to lead to ASI. Coalition members draw on their experience restricting research in other dangerous fields such as nuclear and chemical weapons. These research restrictions aim to be as narrow as possible, preventing the creation of more capable general AI models while still allowing safe and beneficial narrow AI models to be created. The coalition encourages and makes explicit carve-outs for safe, application-specific AI activities, such as self-driving cars and other uses that provide benefits to society.

Verification that members are adhering to research restrictions is aided by the fact that relatively few people have the skills to contribute to this research—likely only thousands or tens of thousands, see Appendix~\ref{app:agreement}, Article IX. Nations verify compliance by way of intelligence gathering, interviews with the researchers, whistleblower programs, and more. This verification aims to use non-invasive measures to ensure that researchers are not working on restricted topics. Additionally, inspectors verify that the medium-scale training allowed by the agreement uses only approved methods (e.g., from a Whitelist) and does not make use of any novel AI methods or algorithms; these would be evidence that restricted research had taken place.

The coalition’s methods for detecting covert violations of the agreement do not prevent overt violations. For these, the coalition relies on nonproliferation and enforcement. \textbf{Nonproliferation efforts focus primarily on states that are not part of the agreement}, beginning with export controls on AI chips and chip manufacturing equipment as well as strong counterproliferation efforts to ensure these export controls are upheld.

Should nonproliferation fail, or if a rogue actor is found to be working towards ASI, coalition members can employ their agreed-upon \textbf{enforcement mechanisms}. These start with standard tools of international pressure such as diplomacy and economic sanctions. Depending on the nature of the developments, countries could escalate to other means at their disposal to disrupt, slow, or impede ASI development. This draws parallels to other contexts where countries have intervened to stop violations of arms-control and nonproliferation agreements.

This international agreement is intended to persist until the conditions surrounding AI development are no longer catastrophically dangerous. Although the exact details remain open to debate, conditions for lifting restrictions may include a strong AI alignment research field with a years-long track record of successfully solving alignment-related problems, an established consensus about the efficacy of ASI-alignment methods, strong misuse and proliferation controls, and the creation of a globally-monitored AI development project that proceeds with due caution. World leaders may also wish to stipulate a requirement to solve or mitigate specific AI-related challenges, including unemployment, power concentration, geopolitical destabilization, and misuse.

While the goal is for this agreement to last as long as is necessary, that may not be practical. There are numerous reasons the agreement might fail; for instance, two threat actors are countries outside the agreement and covert state-backed ASI efforts from countries that are in the agreement. The verification, nonproliferation, and enforcement measures in the agreement are set up with the purpose of preventing such failures, but it’s hard to predict how successful they will be. World leaders may find themselves weighing the risk of a covert project building ASI despite the agreement against the risk of nullifying the agreement and building ASI themselves.

There is not yet political will to implement this agreement, but there might be political will to enact measures that build toward it. Therefore, in Appendix~\ref{app:stages} we describe a staged implementation of this agreement. The staged approach starts with international confidence building and transparency measures, and eventually builds up to this full agreement. A staged approach better mirrors how international arms control agreements have taken place historically, and it allows action to happen today even while the political will for a full agreement does not exist today.

\section{Why this plan in particular?}
\label{sec:whythisplan}

\paragraph{\textit{Wouldn’t a less costly plan work?}}
We have aimed to articulate a solution that could actually prevent the development of ASI. Various parts of this plan, such as restrictions on research and halting for decades, might seem extreme. \textbf{This section discusses why we believe these components are necessary to a successful plan}. We organize the section around objections readers might have, providing our response in turn.

\paragraph{\textit{Would the relevant actors implement such an agreement?}}
As of mid-2025, the U.S. and PRC are not sufficiently concerned about the risks from ASI to agree to halt AI development. But, if people, including world leaders, understood the extreme risks, we believe there would be political will for leaders to take action. Because any actor building ASI, anywhere on earth, would threaten everyone, preventing premature development requires global enforcement. Even a 10\% chance of human extinction should be enough to motivate this.

The implementation of this international agreement can also be staged, as we propose in this paper and detail in Appendix~\ref{app:stages}. This would entail taking less costly actions initially, and then taking more substantial steps as political will increases.

\paragraph{\textit{Why not wait and pause later?}}
The field of AI lacks reliable methods to predict when AI systems will develop dangerous capabilities. Given this fundamental uncertainty, we should set conservative limits on AI development, rather than risk overshooting catastrophic thresholds.

Crucially, we need to stop before dangerous AI systems have been trained. If an AI system has been trained that could cause catastrophic harm via inference, then governance must apply controls to both inference and training. This would also be the case for AI that can substantially automate AI development itself, as this could accelerate development of more efficient algorithms that could compensate for restrictions on hardware. Our current framework focuses on preventing training, but controls on inference would require more invasive measures.

Inaction will also lead to further AI hardware proliferation, making comprehensive tracking and monitoring harder. Additionally, while large secret state-run AI facilities likely don't exist today, this may change as governments recognize the geopolitical implications of ASI. The potential for these secret facilities may undermine the agreement.

\paragraph{\textit{Why are you prohibiting research?}}
We recognize that restricting research is controversial and normally a bad idea. Unfortunately, we think it is necessary to prevent the premature development of ASI. This is primarily a descriptive statement, not a normative one: to prevent the development of ASI, the narrow set of research aimed at this goal must be stopped. In the current paradigm, AI capabilities increase from a combination of scaling computing power and having better AI algorithms; limiting only one of these factors would allow AI capabilities to improve through the other.

The risks from continued research can be decomposed into three types:

First, the fast grind of algorithmic progress: historical trends in AI algorithmic progress indicate that the number of operations used to train an AI to a given capability level drops by 3$\times$ each year \citep{ho_algorithmic_2024}. This means that each year, the threshold size of AI chip clusters that must be monitored to maintain the same level of verification of final AI capabilities would also decrease by 3$\times$. If such trends are not interrupted, widespread consumer computers would eventually become dangerous, and it would be insufficient to monitor data center AI chip use. It is unclear whether consumer computers \textit{even could} be monitored, but regardless of effectiveness, it is certainly not morally desirable to do so.

Second, progress is sometimes much faster than these average historical trends. For instance, Ho et al. 2024 estimate that the transformer architecture provided 7.2$\times$ reduction in operations, or around two years of progress on its own \citep{ho_algorithmic_2024}. Future major innovations, for example, that make much better use of inference or fine-tuning operations, would pose a major risk to the agreement and its verification. These are especially concerning because there could be little warning about them; by contrast, if progress was exactly 3$\times$ per year, there would be plenty of warning signs that the agreement is becoming unstable.

Third, paradigms for developing AIs other than deep learning could bear fruit. These other paradigms are likely to use far fewer computational resources than the current development paradigm, so the agreement would not be robust to their success.

We are not excited about the prospect of restricting research. Unfortunately, this seems necessary. We are hopeful that this research restriction can be relatively narrow and have negligible collateral damage outside of AI, even if it must be somewhat broad within AI.

\paragraph{\textit{Why do we need to halt for so long?}}
In short, \textbf{AI alignment is probably a difficult technical problem, and it is hard to be confident about solutions}. Pausing for a substantial period gives humanity time to be careful in this domain rather than rushing. Pausing for a shorter amount of time (e.g., 5 years) might reduce risk substantially compared to the current race, but it also might not be enough. In general, world leaders should weigh the likelihood and consequence of different risks and benefits against each other for different lengths of a pause.

Section~\ref{sec:overview_of_ai_risks} discusses some of the reasons why the AI alignment problem may be difficult. Generally, experts vary in their estimates of the difficulty of this problem and the likelihood of catastrophe, with some expecting the problem to be very hard \citep{grace2025thousands, controlai_quotes_2025, pdoom2025wikipedia}. Given this uncertainty about how difficult this problem is, we should prepare to pause for a long time, in case more effort is needed. Our agreement would allow for a long halt, even if world leaders \textit{later} came to believe a shorter one was acceptable.

We also contend that there are other problems which need to be addressed during a halt even if one presumes that alignment can be quickly solved, and these problems are also of an uncertain difficulty. These include risks of power concentration, human misuse of AIs, mass unemployment, and many more. World leaders will likely want at least years to understand and address these problems. The international agreement proposed in this paper is primarily motivated by risks from AI misalignment, but there are numerous other risks that it would also help reduce.

\paragraph{\textit{What about automating AI alignment research?}} One objection to this paper’s plan is that a short pause may suffice because AI alignment research could be automated. Automating AI alignment research was the goal of OpenAI’s now-disbanded Superalignment team \citep{leike2023introducing}, and is a key focus of Anthropic’s Alignment Science team \citep{anthropic2025alignment}. AI companies themselves have not publicly provided detailed plans for how they plan to automate alignment research. Assuming they had such a plan, and assuming they acted with much more caution than the AI field of today, this plan \textit{might} work. But we don’t think one should be very confident in the plan’s success, and it would be an irresponsible gamble with immense stakes. As discussed, the state of the AI alignment field is nascent and has made little progress so far. There are various reasons to expect this plan to fail, or at least to not be confident in its success. This topic has been debated at length in internet forums, and for the sake of concision, we will only briefly outline a couple arguments:
\begin{itemize}[nosep]
    \item Alignment problems for ASI appear difficult. Early AGIs may not be able to solve these problems quickly. More capable AGIs would have a better chance of solving these problems, but creating such AIs takes on more catastrophic misalignment risk. \citep{wentworth2025case, lermen2025why}
    \item Early AGIs may not be aligned or controllable. This plan requires that early AGIs either be \textit{aligned themselves} and thus trying to help solve alignment challenges, or that they are \textit{misaligned but we are still able to get useful labor out of them} (as the AI Control research agenda aims to do~\cite{greenblatt_ai_2024}). Each of these conditions seems prima facie unlikely, and it seems difficult to obtain justifiable confidence in at least one of them working.
\end{itemize}

\paragraph{\textit{Why do we have to track down existing AI chips? That seems difficult and the flow of newly produced chips will overwhelm the existing stock quickly.}}

One core goal of our plan is to prevent any data center capable of training ASI from evading monitoring. This will require locating existing AI chips. Some existing plans don’t include locating existing AI chips, while other plans seem to implicitly rely on it, such as for retrofitting chips with verification mechanisms.

Simply monitoring newly produced chips is insufficient for our goals due to the existing chip stock. One might argue that new chips will comprise a much larger fraction of the world’s computing power than existing chips. However, risk derives from the absolute number of unmonitored chips, not just their proportion in the total chip stock.

Voluntary submission of chips for monitoring, while politically easier, can likely not provide the assurance needed. Some actors will inevitably refuse to participate, and even willing participants won't trust each other. PRC will assume the U.S. is hiding chips; the U.S. will assume the same about the PRC. This dynamic could undermine the whole agreement.

Verification mechanisms require physical access to be effective. With current technology, inspectors cannot verify that chips are not being misused without first knowing where they are. Remote attestation and other technical solutions are in early stages and, even if developed, could likely be circumvented if the chips' physical locations remain unknown. For example, chip owners could simply choose not to update their chips’ firmware.

\paragraph{\textit{Can’t we just build defensive technology and adapt? That’s how we manage most dangerous technology.}}
Building defensive technologies and improving societal resilience are noble aims, but, unfortunately, appear insufficient in this case. There are likely to be AI-enabled or AI-accelerated technologies and strategies which are offense-dominant, where offensive use is relatively cheap and that are infeasibly difficult to defend against. Biological weapons are one possible example. Even if a technology is defense-dominant \textit{in the limit,} defense may be \textit{initially} outpaced by offensive uses. For example, it may eventually be possible to harden all critical cyber infrastructure against attacks, but there will likely be a window where there are AI systems capable of performing attacks before AI systems have hardened all such infrastructure. Additionally, ASI may enable entirely new classes of technology, such as advanced nanotechnology or novel methods of psychological manipulation. It seems overwhelmingly likely there will be at least one instance of a new technology that is catastrophically dangerous and too difficult to defend against.

\paragraph{\textit{Won’t this plan enable authoritarianism?}} This agreement does include elements that could be misused for authoritarian abuse, such as centralizing AI infrastructure, monitoring training runs, and tracking researchers. \textbf{However, the default trajectory carries far greater authoritarianism risk.} ASI may be highly power-concentrating: if someone manages to develop an aligned superintelligence, it may grant them unrivalled economic, military, or technological power. Under current conditions, ASI will likely be built by a private company (potentially in partnership with government agencies), with minimal public oversight. In this situation, control over the most powerful technology would rest with perhaps 1-10 individuals. This would be an unprecedented concentration of power with few mechanisms to prevent permanent authoritarian lock-in. By contrast, this agreement prevents the creation of ASI, allowing time to establish the required governance structures.

The agreement's verification mechanisms (e.g., tracking chips, monitoring training runs, limiting research) would likely involve a combination of existing and new authorities. As a few examples of related precedent, the U.S. government already conducts extensive surveillance (including against U.S. persons), maintains export controls on sensitive technologies, inspects nuclear facilities, and restricts the publication of nuclear research. Where domestic implementation expands the government’s power, there are likely safeguards that could be implemented to prevent misuse. For instance, multilateral controls, narrow scoping of authority, and transparency measures would reduce these risks.

\paragraph{\textit{Doesn’t this plan take on too much risk by being slow to implement and allowing chips to continue existing?}}
This plan aims to strike a balance between political viability and efficacy, balancing costs and benefits, but it may not balance them optimally. One concern is that the existence of AI chips makes breakout time too short: if some country decided to flout the agreement and attempt to build ASI, they might be able to do so quickly, leaving little time for others to thwart them. By contrast, if existing chips and the chip supply chain were decommissioned, it might take a decade or more for a country to build the necessary AI infrastructure, making it much easier for others to intervene. We are optimistic that strong verification tools accompanied by political will for enforcement would be sufficient to prevent rogue states from building ASI, even when AI infrastructure is allowed to exist.

Another concern relating to the plan is that it relies too heavily on remaining in the current AI development paradigm where significant computational resources are needed to improve AI capabilities. There are early signs that this may be changing, such as the inference-scaling regime started with OpenAI’s o-series of models~\citep{jaech2024openai}. If implemented soon, this agreement would likely be sufficient, but the technical situation could deteriorate if action is deferred. Overall, our choice to go with this more permissive agreement—one that allows existing chips to keep existing and focuses primarily on preventing large training—is an attempt to balance costs and benefits, trying to preserve the safe use of existing AI models without the risks from continued capabilities advancement.

\section{Conclusion}

The development of artificial superintelligence represents an immense challenge for human civilization. Unlike most technological risks, where defensive measures and gradual adaptation can provide adequate protection, ASI poses an existential threat that demands preemptive action. This paper has outlined an international agreement centered on halting dangerous AI capabilities advancement through FLOP-based training restrictions, comprehensive AI chip tracking, and targeted research restrictions. While the political will for such an agreement does not exist today, the stakes are too high to accept inaction. A 10\% chance of catastrophic outcomes is considered unacceptable in any other domain of national security or public safety.

We recognize that this proposal faces significant obstacles: it requires unprecedented international cooperation, restrictions on the use of AI chips, and limitations on scientific research. However, we believe these measures are proportionate to the risks and represent the most viable path to ensuring that ASI, when eventually developed, will be beneficial for humanity. The staged implementation in the appendices provides a pathway forward even in the absence of immediate political consensus.

As AI capabilities continue their rapid advancement, the window for establishing robust verification and monitoring regimes narrows. We hope this work can serve as a foundation for serious policy discussions and international negotiations, guiding humanity toward a future where the benefits of advanced AI can be realized without courting civilizational catastrophe.

\section*{Acknowledgments}
We are thankful to the following people and numerous others for feedback on earlier versions of this work. They do not necessarily endorse the views expressed here.

Nate Soares, Mitch Howe, Joseph Rogero, Robi Rahman, Rob Bensinger, Malo Bourgon, Daniel Kokotajlo, Thomas Larsen, Oscar Delaney, Naci Cankaya.

LLMs were used to help with various parts of this report, including background research, text drafting, and text refinement. All final text was reviewed by a human and the vast majority was written by a human.

\bibliographystyle{plainnat}
\bibliography{references} 

\newpage

\appendix

%

\section{The Agreement}
\label{app:agreement}
Below, we provide an annotated example draft language for the sort of agreement that could be implemented by major governments around the world, if they recognized the dangers from artificial superintelligence (ASI) and sought to prevent anyone from building ASI. We present this as an illustrative example of some potentially valuable provisions to have in view, using mechanisms tailored to the situation at hand and grounded in historical precedent.

This draft text covers many different mechanisms that we think would be required to prevent AI developers from seriously endangering humanity. In practice, we would expect different aspects to likely be covered by different agreements.\footnote{This is the case with nuclear weapons agreements, where separate treaties establish the IAEA (\href{https://www.iaea.org/about/overview/statute}{1956}, by the Conference on the Statute of the International Atomic Energy Agency, hosted at the Headquarters of the United Nations), the NPT (\href{https://www.iaea.org/sites/default/files/publications/documents/infcircs/1970/infcirc140.pdf}{1970}, through negotiations in the United Nations Eighteen Nation Committee on Disarmament), and the arms control agreements like the START treaty (\href{https://media.nti.org/documents/start_1_treaty.pdf}{1991}, following nine years of intermittent negotiation between the U.S. and the Soviet Union).} And of course, in reality, the parties involved should draft the agreement subject to negotiation and review by relevant experts.

For each article, we’ve provided a commentary section explaining why we made key decisions, and a section discussing some relevant precedent. It's important to note that precedent is valuable because it demonstrates that the plan can be effectuated, not that it will be effective. The existence of prior agreements shows that similar mechanisms are politically and practically feasible, but does not guarantee they will achieve their intended outcomes in this domain. By contrast, our reasons for believing this particular agreement would be effective are discussed in various places, including the commentary sections, and are often AI-specific.

A real agreement would involve many details. We’ve included some level of detail, but also relegate much detail to “annexes” which would have to be finalized later. Many of the quantities and numerical thresholds we use in our draft constitute our best guess, but they should still be treated only as guesses. Many of those numbers would require further study and revision before being finalized. These sorts of details plausibly wouldn’t be included in the agreement itself, analogous to how, in the case of the Treaty on the Non-Proliferation of Nuclear Weapons (NPT), specific details of inspections and so-called “safeguards” programs were decided between each country and the IAEA, rather than being included in the NPT itself. However, for clarity, we have kept our best-guess numbers directly in the text, to help it feel more concrete.

\subsection*{Preamble}

The States concluding this Agreement, hereinafter referred to as the Parties to the Agreement,

Alarmed by the prospect that the development of artificial superintelligence would lead to the deaths of all people and the end to all human endeavor,

Affirming the necessity of urgent, coordinated, and sustained international action to prevent the creation and deployment of artificial superintelligence under present conditions,

Convinced that the measures to prevent advancement of artificial intelligence capabilities will reduce the chance of human extinction,

Recognizing that the stability of this Agreement relies on the ability to verify the compliance of all Parties,

Recalling the precedent of prior arms control and nonproliferation agreements in addressing global security threats,

Undertaking to cooperate in facilitating the verification of artificial intelligence activities globally when they steer well clear of artificial superintelligence, and seeking to preserve access to the benefits of artificial intelligence systems even while avoiding dangers,

Have agreed as follows:

\paragraph{Precedent for this Preamble}
The preamble of this agreement is modeled after that of The Treaty on the Non-Proliferation of Nuclear Weapons (NPT), which begins:

\begin{displayquote}
Considering the devastation that would be visited upon all mankind by a nuclear war and the consequent need to make every effort to avert the danger of such a war and to take measures to safeguard the security of peoples…
\end{displayquote}

and soon adds:

\begin{displayquote}
Affirming the principle that the benefits of peaceful applications of nuclear technology, including any technological by-products which may be derived by nuclear-weapon States from the development of nuclear explosive devices,

should be available for peaceful purposes to all Parties to the Treaty, whether nuclear-weapon or non-nuclear-weapon States…
\end{displayquote}

In so doing, the preamble invites the world to join responsible parties in safeguarding humanity from the catastrophic threat of a powerful technology, and to share in the benefits that can be safely permitted. Our preamble tries to follow this example.

The NPT entered into force in 1970 and was extended indefinitely in 1995. Known for its near-universal membership (191 parties), its preamble emphasizes the global hazard of weapons proliferation while affirming that the benefits of peaceful nuclear applications should be available to all parties.

\subsection*{ARTICLE I — Primary Purpose}

Each Party to this Agreement does not develop, deploy, or seek to develop or deploy artificial superintelligence ("ASI") by any means. Each Party prohibits and prevents all such development within their borders and jurisdictions, and, due to the uncertainty as to when further progress would produce ASI, does not engage in or permit activities that materially advance toward ASI as described in this Agreement. Each Party assists, or does not impede, reasonable measures by other Parties to dissuade and prevent such development by and within non-Party states and jurisdictions. Each Party implements and carries out all other obligations, measures, and verification arrangements set forth in this Agreement.

Where some classes of AI infrastructure and capabilities staying far from ASI may be deemed acceptable but only under conditions of international supervision, only Parties to the Agreement may carry out such activities, or own or operate AI chips and manufacturing capabilities that could potentially lead to the development of ASI if unsupervised. Non-Parties are denied such access for the safety of the Parties and of all life on Earth (Article V, Article VI, Article VII).

Parties commit to a dispute resolution process (Article XI) to minimize unnecessary Protective Actions (Article XII).

\paragraph{Precedent for Article I}
Article I of the NPT, as in many treaties, states the high-level commitment parties are making — in this case, to not share their nuclear weapons or help others obtain them:

\begin{displayquote}
    Each nuclear-weapon State Party to the Treaty undertakes not to transfer to any recipient whatsoever nuclear weapons or other nuclear explosive devices or control over such weapons or explosive devices directly, or indirectly; and not in any way to assist, encourage, or induce any non-nuclear-weapon State to manufacture or otherwise acquire nuclear weapons or other nuclear explosive devices, or control over such weapons or explosive devices.
\end{displayquote}

The commitment summarized in Article I of our draft agreement is stronger than this because an ASI breakout by anyone, anywhere, cannot be allowed to happen even once.\footnote{The NPT is generally credited with keeping the number of nuclear states lower than it might have been, but acquisitions by non-signatories (India, Pakistan, Israel) and former signatories (North Korea) have still occurred. Any non-signatory creating even a single ASI is comparable in danger to a mass thermonuclear exchange, and must be treated accordingly.} It would not be enough to not “assist, encourage, or induce” others to build it. We have therefore included a commitment to “assist, or not impede, reasonable measures” by parties to dissuade and prevent such development anywhere.

The NPT works to contain an existing threat (nuclear weapons), while our draft agreement is working to prevent a threat from existing at all (ASI). Precedent for preventing the development of dangerous new technology can be found in the Protocol on Blinding Laser Weapons, part of the Convention on Certain Conventional Weapons\footnote{The Convention on Prohibitions or Restrictions on the Use of Certain Conventional Weapons Which May Be Deemed to Be Excessively Injurious or to Have Indiscriminate Effects, commonly called the CCW, entered into force in 1983. As of 2024, its 128 parties commit to protect combatants and non-combatants from unnecessary and egregious suffering by restricting various categories of weapons.}. Its Article I reads:

\begin{displayquote}
It is prohibited to employ laser weapons specifically designed, as their sole combat function or as one of their combat functions, to cause permanent blindness to unenhanced vision, that is to the naked eye or to the eye with corrective eyesight devices. The High Contracting Parties shall not transfer such weapons to any State or non-State entity.
\end{displayquote}

That language doesn’t try to keep anyone anywhere from ever testing or accidentally making such a system, however. Our agreement must be strong enough to prevent ASI from being made accidentally. Because it’s not clear where the point-of-no-return might be, our Article I includes a commitment to “not engage in or permit activities that materially advance toward ASI.”

\subsection*{ARTICLE II — Definitions}

For the purposes of this Agreement:

\begin{enumerate}
    \item \textbf{Artificial intelligence (AI)} means a computational system that performs tasks requiring cognition, planning, learning, or taking actions in physical, social or cyber domains. This includes systems that perform tasks under varying and unpredictable conditions, or that can learn from experience and improve performance.
    \item \textbf{Artificial superintelligence (ASI)} is operationally defined as any AI with sufficiently superhuman cognitive performance that it could plan and successfully execute the destruction of humanity.
    \begin{enumerate}
        \item For the purposes of this Agreement, AI development which is not explicitly authorized by the Coalition Technical Body (Article III) and is in violation of the limits described in Article IV shall be assumed to have the aim of creating artificial superintelligence.
    \end{enumerate}
    \item \textbf{Dangerous AI activities} are those activities which substantially increase the risk of an artificial superintelligence being created, and are not limited to the final step of developing an ASI but also include precursor steps as laid out in this Agreement. The full scope of dangerous AI activities is concretized by Articles IV through IX and may be elaborated and modified through the operation of the Agreement and the activities of the Coalition Technical Body.
    \item \textbf{Floating-point operations (FLOP)} is the computational measure used to quantify the scale of training and post‑training, based on the number of mathematical operations done. FLOP shall be counted as either the equivalent operations to the half-precision floating-point (FP16) format or the total operations (in the format used), whichever is higher.
    \item \textbf{Training run} means any computational process that optimizes an AI’s parameters (specifications of the propagation of information through a neural network, e.g., weights and biases) using gradient-based or other search/learning methods, including pre-training, fine-tuning, reinforcement learning, large-scale hyperparameter searches that update parameters, and iterative self-play or curriculum training.
    \item \textbf{Pre-training} means the training run by which an AI’s parameters are initially optimized using large-scale datasets to learn generalizable patterns or representations prior to any task- or domain-specific adaptation. It includes supervised, unsupervised, self-supervised, and reinforcement-based optimization when performed before such adaptation.
    \item \textbf{Post-training} means a training run executed after a model’s pre-training. In addition, any training performed on an AI created before this Agreement entered into force is considered post-training.
    \item \textbf{Strict Threshold} is the amount of training computation (measured in FLOP) above which training runs are prohibited. It is set at $10^{24}$ FLOP.
    \item \textbf{Strict Post-training Threshold} is the amount of training computation (measured in FLOP) above which post-training runs (e.g., of models trained before the agreement) are prohibited. It is set at $10^{23}$ FLOP.
    \item \textbf{Monitored Threshold} is the amount of training computation (measured in FLOP) above which training runs are subject to monitoring by the international authority. It is set at $10^{22}$ FLOP.
    \item \textbf{Advanced computer chips} are integrated circuits fabricated on processes at least as advanced as the 28 nanometer process node.
    \item \textbf{AI chips} mean specialized integrated circuits designed primarily for AI computations, including but not limited to training and inference operations for machine learning models [this would need to be defined more precisely in an Annex]. This includes GPUs, TPUs, NPUs, and other AI accelerators. This may also include hardware that was not originally designed for AI uses but can be effectively repurposed. AI chips are a subset of advanced computer chips.
    \item \textbf{AI hardware} means all computer hardware for training and running AIs. This includes AI chips, as well as networking equipment, power supplies, and cooling equipment.
    \item \textbf{AI chip manufacturing equipment} means equipment used to fabricate, test, assemble, or package AI chips, including but not limited to lithography, deposition, etch, metrology, test, and advanced-packaging equipment [a more complete list would need to be defined in an Annex].
    \item \textbf{H100-equivalent} means the unit of computing capacity (FLOP per second) equal to one NVIDIA H100 SXM accelerator, 989 TFLOP/s in FP16, or a Total Processing Performance (TPP) of 15,824 TFLOP-bit/s, where TPP is calculated as TPP = 2 $\times$ non-sparse MacTOPS $\times$ (bit length of the multiply input).
    \item \textbf{Covered chip cluster (CCC)} means any set of AI chips or networked cluster with aggregate effective computing capacity or accelerator memory greater than 16 H100-equivalents. A networked cluster refers to chips that either are physically co-located, have inter-node aggregate bandwidth — defined as the sum of bandwidth between distinct hosts/chassis — greater than 25 Gbit/s, or are networked to perform workloads together. The aggregate effective computing capacity of 16 H100 chips is 15,824 TFLOP/s, or TPP of 253,184 TFLOP-bit/s, and is based on the sum of per-chip TPP. The total accelerator memory of 16 H100 chips is 1,280 GB. Examples of CCCs would include: the GB200 NVL72 server, three eight-way H100 HGX servers residing in the same building, CloudMatrix 384, a pod with 32 TPUv6e chips, every supercomputer.
    \item \textbf{National Technical Means (NTM)} includes satellite, aerial, cyber, signals, imagery (including thermal), and other remote-sensing capabilities employed by Parties for verification consistent with this Agreement.
    \item \textbf{Chip-use verification} means methods that provide insight into what activities are being run on particular computer chips in order to differentiate acceptable and prohibited activities.
    \item \textbf{Methods used to create frontier models} refers to the broad set of methods used in AI development. It includes but is not limited to AI architectures, optimizers, tokenizer methods, data curation, data generation, parallelism strategies, training algorithms (e.g., RL algorithms) and other training methods. This includes post-training but does not include methods that do not change the parameters of a trained model, such as prompting. New methods may be created in the future.
    \item \textbf{AI Technique Whitelist} means the list of approved AI methods and techniques maintained by the Coalition Technical Body. Training runs above the Monitored Threshold may only use techniques on this list.
\end{enumerate}

\paragraph{Notes on Article II}

\subparagraph{On Definitions of AI}
The definition of AI used here (adapted from Senator Chuck Grassley’s \href{https://www.congress.gov/bill/119th-congress/senate-bill/1792/text}{AI Whistleblower Protection Act}) is possibly too broad. Further refinement would help make it clear that the definition should not apply to obviously-safe computer systems such as spellcheck or image recognition systems.

If AI technology were never going to change from its modern form, in which development for a frontier Large Language Model requires highly specialized hardware and is easily distinguishable from other activities, it would be easier to craft a narrow tailored definition. But ASI is a moving target, and the definition of AI that is used must cover more than just LLMs. An agreement prohibiting solely machine learning might encourage researchers to develop new AI paradigms that don’t technically meet the definitions, so that they can race ahead toward superintelligence. If a novel paradigm did emerge, especially one which is not as AI-chip-intensive as deep learning, then the agreement would likely need to be updated, and enforcement might become substantially more difficult.

\subparagraph{On Definitions of Computing Capacity}
We use H100-equivalent as the primary metric for computing capacity. In Article V, this is used to set the size of the largest allowed unmonitored chip cluster (16 H100-equivalents). Article IV defines thresholds in terms of the total operations used to train an AI, and so, by setting limits on unmonitored operations per second, this effectively would make it infeasibly slow to conduct an illegally large training run on unmonitored hardware.

We use H100-equivalents because the most relevant metric in various chip designs is how quickly they perform operations, and H100s serve as a fine and precedented measuring stick. Other chip metrics are important in AI training (such as high bandwidth memory), but overall, these matter less than the number of operations per second.

Our proposed definition of a covered chip cluster (CCC) is an attempt to satisfy several constraints: The bound should be high enough to prevent regular people from breaking the rules (i.e., 25 Gbit/s bandwidth between chassis is faster than non-data center internet connections; it is very rare and expensive for an individual to own more than 16 H100-equivalents). The bound must also be set low enough to prevent dangerous AI activities and to make subversion difficult (i.e., make it difficult to do training distributed across multiple sub-CCC sets of chips). We discuss the tradeoffs more in the notes after Article V. The constraints on CCC accelerator memory are designed to prevent illicit distributed training, \href{https://techgov.intelligence.org/blog/catching-illicit-distributed-training-operations-during-an-ai-pause}{as discussed in this article}. 

AI chips are a subset of advanced computer chips, and there isn’t a bright line that distinguishes AI chips from non-AI chips. Instead of defining and relying on a distinction here, we use the overall computing capacity (in operations per second) of a cluster, as measured in H100-equivalents. If the chips could be configured for training or running AIs and are above the defined threshold, then the agreement requires that they be monitored.

Note that National Technical Means (NTM) may be deprecated as the official term by some governments. We use it in this agreement in the style of past arms control agreements for ease of comparison.

\subsection*{ARTICLE III — The Coalition}

\begin{enumerate}
    \item Parties to this Agreement constitute the coalition. The coalition shall implement this Agreement and its provisions, including those for international verification of compliance with it, and shall provide a forum for consultation and cooperation among Parties.
    \item The organs of the coalition are the Executive Council and the Coalition Technical Body (CTB).
    \item Executive Council
    \begin{enumerate}
        \item The Executive Council initially consists of the United States of America and the People's Republic of China.
        \item The Executive Council: approves challenge inspections; appoints the Director-General; provides oversight of the CTB and exercises veto power over its recommendations; determines overall policy and adopts the budget.
        \item Decision-making processes are as follows:
        \begin{enumerate}
            \item All proactive Executive Council decisions require consensus among members. If consensus cannot be reached, the proposed changes are not adopted.
            \item Each member of the Executive Council has veto power over decisions by the CTB.
            \item The Executive Council may delegate specific authorities to the CTB, subject to the veto power described above.
        \end{enumerate}
        \item The Executive Council may hold deliberative sessions with selected additional Parties that are not on the Executive Council. These additional Parties are chosen by the Executive Council. Selected Parties may participate in debate and be provided with relevant sensitive information in order to do so.
    \end{enumerate}
    \item Coalition Technical Body (CTB) and Director-General
    \begin{enumerate}
        \item The Director-General of the CTB is its head and chief administrative officer.
        \item The Director-General is appointed by the Executive Council for a four-year term, renewable once. The Executive Council can recall the Director-General.
        \item The CTB coordinates the activities of the Parties required by the Agreement. It includes technical divisions for Chip Tracking and Manufacturing Safeguards, Chip Use Verification Safeguards, Research Controls, Information Consolidation, Technical Reviews, Administration and Finance, and Legal and Compliance. The Director-General can create and disband technical divisions.
        \item The CTB, through the Director-General, proposes changes to technical definitions and safeguard protocols, as necessary to implement Article IV, Article V, Article VI, Article VII, Article VIII, Article IX, and Article X of this Agreement.
        \begin{enumerate}
            \item Time-sensitive changes to FLOP thresholds (Article IV), the size of covered compute clusters (Article V), and the boundaries of restricted research (Article VIII) may be implemented by the Director-General immediately in the case where inaction poses a security risk. Such changes remain in effect for thirty days. Past that, the changes require approval from the Executive Council to remain in effect, subject to the veto power of each Executive Council member.
        \end{enumerate}
    \end{enumerate}
    \item The coalition's regular budget is funded by assessed contributions of members of the Executive Council, with the assessment scale determined by the Executive Council.
\end{enumerate}

\paragraph{Precedent for Article III}
The \href{https://2009-2017.state.gov/t/avc/trty/102360.htm#text}{Intermediate-Range Nuclear Forces} (INF) Treaty and Strategic Arms Reduction Treaties (\href{https://1997-2001.state.gov/www/global/arms/starthtm/start/start1.html}{START I}, \href{https://2009-2017.state.gov/t/avc/trty/104150.htm}{START II}, and \href{https://2009-2017.state.gov/documents/organization/140035.pdf}{New START}), place responsibility for implementation and verification on the individual parties; each commit to procedures that allow the other to obtain reasonable assurance of compliance.

The Executive Council established by paragraph 3 emulates the NPT’s Board of Governors.

Taiwan complicates our agreement concept, given its delicate geopolitical situation and its status as the producer of most of the world’s AI chips. Fortunately, precedent provides guidance: Though Taiwan is not a party to the NPT, it has stated on multiple occasions that it considers itself bound by the principles of the NPT. Taiwan allows the IAEA to conduct inspections and apply safeguards to its nuclear facilities through a trilateral agreement with the United States and the IAEA. A similar arrangement could be worked out with regard to this agreement.

The “challenge inspections” in paragraph 3, subparagraph (b) are modeled after the mechanism in Part X of the CWC; we will elaborate on this precedent with Article X.

\paragraph{Notes on Article III}
This arrangement centralizes only those few functions which must be centralized (such as maintaining and clarifying limits on AI research, development, and deployment) within the Coalition Technical Body and provides oversight power to the Executive Council. One benefit to our draft structure is that it enables the technical body to carry out rapid decision-making, though these decisions must survive being vetoed by any Executive Council member.

In this agreement, the U.S. and PRC, and possibly other executive council members pursue parallel efforts at verifying and enforcing compliance (see also Article X). This aims to meet each party’s assurance needs while sacrificing the smallest amount of autonomy and not at all hampering pre-existing efforts at intelligence gathering.

It should be noted that the U.S. and PRC are the preeminent powers when it comes to AI technology. The viability of any agreement would require that both countries are party to any agreement whether it be bilateral or otherwise. Realistically, both countries will have an outsized say over restrictions, supply, and research given the current state of AI development. With this in mind, it’s also important to acknowledge that certain states, such as the permanent members of the UN Security Council, along with countries with substantial economic and military resources will only join this agreement if it is in line with their national security and economic interests. We are writing this agreement in such a way as to not take a preemptive stand on which countries would join the Executive Council. Ultimately it will be up to interested parties to the agreement to negotiate who is in the Executive Council based on relative negotiation advantages and positions going into such discussions.

Given the status of TSMC as the leading AI chip manufacturer, any AI agreement must consider how to address Taiwan. As discussed in the Precedent section, we would encourage Taiwan to adhere to our agreement much as it adheres to the NPT without having signed it, through formal arrangements and/or declarations stating that Taiwan considers itself to be bound by the principles of this agreement and is open to on-site routine and/or challenge inspections.

\subsection*{ARTICLE IV — AI Training}

\begin{enumerate}
    \item Each Party agrees to ban and prohibit AI training above the following thresholds: Any training run exceeding the Strict Threshold or any post-training run exceeding the Strict Post-training Threshold. Each Party agrees to not conduct training runs above these thresholds, and to not permit any entity within its jurisdiction to conduct training runs above these thresholds.
    \begin{enumerate}
        \item The Coalition Technical Body (CTB) may modify these thresholds, in accordance with the process described in Article III.
    \end{enumerate}
    \item Each Party shall report any training run above the Monitored Threshold to the CTB, prior to initiation. This applies for training runs conducted by the Party or any entity within its jurisdiction.
    \begin{enumerate}
        \item This report must include, but is not limited to, all training code, all training data, and an estimate of the total FLOP to be used. The Party must provide CTB staff supervised access to all data, with access logging appropriate to the data's sensitivity, and protections against duplication or unauthorized disclosure. Data and code may be reviewed on-site at the training facility or through secure mechanisms approved by the CTB. Failure to provide CTB staff sufficient access to data is grounds for denying the training run, at the CTB's discretion. The CTB may request any additional documentation relating to the training run. The CTB will also pre-approve a set of small modifications that could be made to the training procedure during training. Any such changes will be reported to the CTB when and if they are made.
        \item Training runs above the Monitored Threshold require explicit approval from inspectors representing the U.S. and the PRC, physically present at the training facility. Inspectors must provide written approval before training may commence. Either may withhold approval or require modifications. Inspectors shall verify that the training code uses only AI techniques approved on the AI Technique Whitelist maintained by the CTB.
        \item Inspectors may not remove electronic storage devices or data from the training facility.
        \item The CTB may monitor such training runs, and the Party will provide checkpoints of the model to the CTB upon request, including the final trained model [initial details for such monitoring would need to be described in an Annex].
        \item In the event that monitoring indicates worrisome AI capabilities or behaviors, the CTB can issue an order to pause a training run or class of training runs until it deems it safe for the training run to proceed.
        \item The CTB will maintain robust security practices. The CTB will share information about declared training runs with the U.S. and PRC to support independent verification efforts.
        \item In the event that a Party discovers a training run above the designated thresholds, whether through the Party's own verification efforts or otherwise, the Party must report this training run to the CTB and halt this training run (if it is ongoing). Such a training run may only resume with approval from the CTB.
    \end{enumerate}
    \item Each Party, and entities within its jurisdiction, may conduct training runs of less FLOP than the Monitored Threshold without oversight or approval from the CTB.
    \item The CTB may authorize specific carveouts for activities such as safety evaluations, self-driving vehicles, medical technology, and other activities deemed safe by the Director-General, subject to the Executive Council's veto power under Article III. These carveouts may allow for training runs larger than the Strict Threshold with CTB oversight, or a presumption of approval from the CTB for training runs between the Monitored Threshold and Strict Threshold.
    \item The CTB creates and maintains an AI Technique Whitelist specifying allowed AI methods and techniques. The CTB may modify this Whitelist in accordance with Article III. Training runs above the Monitored Threshold may only employ techniques on this Whitelist.
\end{enumerate}

\paragraph{Precedent for Article IV}
While the numerical values for thresholds specified in our agreement can and should be revisited when moving beyond the early draft stage, quantitative caps are common in international agreements, preempting disputes that would otherwise hinge on differing interpretations of qualitative language.

The 1974 \href{https://2009-2017.state.gov/t/isn/5204.htm}{Threshold Test Ban Treaty} established a cap of 150 kilotons on underground nuclear tests performed by the U.S. and USSR.\footnote{The U.S. and USSR had already agreed to stop other kinds of nuclear weapons tests in 1963 with the Treaty Banning Nuclear Weapon Tests in the Atmosphere, in Outer Space and Under Water, commonly called the \href{https://2009-2017.state.gov/t/avc/trty/199116.htm}{Limited Test Ban Treaty} (LTBT) or Test Ban Treaty.} The purpose and effect of this treaty was to at least somewhat hinder further development of larger and more destructive “city buster” warheads. A relevant parallel to AI development is that, as of mid-2025, more general and capable — and therefore more hazardous — models take correspondingly larger training runs to create; our agreement specifies caps intended to prevent such AIs from being intentionally developed, but also to reap the essential (if non-parallel) benefit of reducing the risk of an unforeseen capabilities threshold being accidentally and irretrievably crossed.

The training limit we have suggested as a starting point is low enough that some AI models trained today would exceed it; we see this as prudent in expectation of advances that make newer models more capable per unit of training (discussed with Article VIII). Arms reduction agreements provide precedent for thresholds set below the current maximum level. The 1922 \href{https://www.digitalhistory.uh.edu/disp_textbook.cfm?smtID=3&psid=3995}{Washington (Naval) Treaty} set warship displacement limits that required the U.S. and other naval powers to scrap dozens of capital ships.\footnote{The \href{https://treaties.fcdo.gov.uk/data/Library2/pdf/1924-TS0005.pdf}{Treaty Between the British Empire, France, Italy, Japan, and the United States of America for the Limitation of Naval Armament} (the Washington Naval Treaty) lists ships to be scrapped by name in a table (Section II).} In Article II of the 1991 START treaty,\footnote{\href{https://1997-2001.state.gov/www/global/arms/starthtm/start/start1.html}{The Strategic Arms Reduction Treaty} was signed in 1991 and entered force in 1994. Signatories were each barred from deploying more than 6,000 nuclear warheads on a total of 1,600 intercontinental ballistic missiles and bombers.} the U.S. and the Soviet Union (and later, the Russian Federation) agreed to limits in the sizes of their nuclear stockpiles and delivery systems that required them to phase out \href{https://bush41library.tamu.edu/archives/public-papers/3263}{more than four thousand warheads each}.

Precedent for quantitative thresholds that limit breakout potential will be discussed with Article V.

\paragraph{Notes on Article IV}
In recent years, advances in AI have followed first and foremost from an increase in computational resources poured into AI training. Restricting these resources, and restricting algorithmic progress research (described in Article VIII), would dramatically reduce the risk that superintelligence could be created in the near term.

The restrictions in our draft are based on the number of computational operations used, as this is relatively easy to define and measure. The performance of existing state-of-the-art AI informs amounts of computing hardware that appear safe, at least when using AI algorithms from mid-2025.

We would prefer to limit training based on the capabilities of the trained system. But no one has the technical ability to confidently predict what a new AI will or won’t be able to do before it is trained. Computational resources are an available proxy.

The hard prohibition at the Strict Threshold ($10^{24}$ FLOP) for training is slightly below that used to train models near the state of the art as of August 2025 (such as DeepSeek-V3, trained with $3\times{10^{24}}$ FLOP). We suggest this threshold because it is below the level at which we expect AIs to be dangerous (given current algorithms), and because it provides some breathing room and a buffer against algorithmic progress.

The prohibition of post-training over $10^{23}$ is meant to apply to the post-training of AIs created prior to the agreement entering into force. Many of these AIs will have been trained using more than $10^{24}$ FLOP; as of mid-2025, there are between 50 and 100 such models. Given that the weights of many such AIs will have been openly released, it is not feasible to prevent people from using them, but it is feasible to prevent large modifications to them via post-training.

A $10^{22}$ FLOP (Monitored Threshold) training run on 16 H100s would take around one week. This is computing-intensive enough that hobbyists would not accidentally break the threshold by training small and allowed models. AIs trained at the $10^{22}$ scale (with current algorithms) have so far appeared to be innocuous, though that situation would need to be monitored, as it could change as a result of unprevented AI research progress.

Monitoring of medium-scale training would allow the CTB to stay (somewhat) up to date with any algorithmic progress that happens in spite of the bans, and Article XIII provides for evaluations of models trained in this band. This would help to apprise the CTB of trends in AI development and adjust the FLOP thresholds however merited.

CTB staff are permitted access to training data used in monitored training runs, subject to a variety of restrictions. The intent of the restrictions is to guarantee that logging and other oversight methods are used to prevent unauthorized disclosure of sensitive contents in the training data, including but not limited to personal identifying information, personal health information, classified data, trade secrets, and banking data subject to secrecy laws.

\subsection*{ARTICLE V — Chip Consolidation}

\begin{enumerate}
    \item Each Party ensures that within their jurisdiction, all covered chip clusters (CCCs), as defined in Article II (i.e., a set of chips with capacity greater than 16 H100-equivalents) [note that 16 H100s collectively cost around \$500,000 in 2025 and these are rarely owned by individuals], are located in facilities declared to the CTB, and that these AI chips are subject to monitoring by the Parties, coordinated by the CTB.
    \begin{enumerate}
        \item Parties aim to avoid co-locating AI chips with non-ancillary non-AI computer hardware in these declared facilities.
        \item These facilities are accessible to physical inspection. This may include, for instance, that verification teams can reach any CCC from at least one airport with scheduled international service within 12 hours.
        \item Parties do not house AI chips in so many different locations that it is infeasible to monitor all locations. If requested by the CTB, Parties must further consolidate their AI chips into fewer monitored facilities.
    \end{enumerate}
    \item Unmonitored AI chips that are not part of a CCC (i.e., that have capacity less than 16 H100‑equivalents) may remain outside of CTB‑declared facilities, provided that such stockpiles are not aggregated or networked to meet the CCC definition, are not rotated among sites to defeat monitoring, and are not used for prohibited training. Parties will make reasonable efforts to monitor the sale and aggregation of AI chips to ensure that any newly created CCCs are detected and monitored and are not used for prohibited training.
    \item Within 120 days of the Agreement entering into force, each Party locates, inventories, and consolidates all CCCs into facilities declared to the CTB. Parties do not disaggregate, conceal, or otherwise reassign chips to evade this requirement or to cause a set of chips which would have been classified as a CCC to no longer be classified as a CCC.
    \item Parties to the Agreement monitor the domestic consolidation process, coordinated by the CTB, including through on‑site inspections, document and inventory verification, accompaniment of domestic authorities during transfers and inspection, and information sharing with Parties under Article X. The CTB may require chain‑of‑custody records for transfers. Parties may conduct challenge inspections as described in Article X. Parties provide timely access to relevant facilities, transport hubs, and records to inspectors conducting monitoring activities. Whistleblower protections and incentives under Article X apply to the consolidation process, and the CTB maintains protected reporting channels.
    \item Within 120 days of the Agreement entering into force, Parties submit to the CTB a register of their CCCs. The register must include the location, type, quantity, serial or other unique identifiers where available, and associated interconnects of all AI chips in the CCCs. Each Party provides the CTB with an updated and accurate register no later than every 90 days.
    \item Parties provide the CTB with advance notice of any planned transfer of AI chips, whether domestic or international, no less than 14 days before the planned transfer. The CTB must approve any transfer before it proceeds. Inspectors are afforded the opportunity to observe the transfer. For international transfers, both the sending and receiving Parties coordinate with the CTB on routing, custody, and receipt. Emergency transfers undertaken for safety or security reasons notify the CTB and Executive Council as soon as practicable, and the CTB will verify the transfer occurred as reported.
    \item Broken, defective, surplus, or otherwise decommissioned AI chips continue to be treated as functional chips, until the CTB coordinates certification of their destruction. Parties do not destroy AI chips without oversight. Destruction or rendering permanently inoperable is conducted under oversight using CTB‑approved methods and recorded in a destruction certificate submitted to the CTB [the details will need to be explained in an Annex]. Salvage or resale of components from such hardware is prohibited unless expressly authorized by the CTB.
\end{enumerate}

\paragraph{Precedent for Article V}
Declaring assets of concern is often a first step in restrictive treaties. Parties to the 1922 Washington Naval Treaty provided inventories of capital ships and their tonnage, and committed to notify each other when replacing these vessels. The 1991 START I treaty included a classified Agreement on Exchange of Coordinates and Site Diagrams (in Article VIII), outlining the sharing of data on the location of all declared strategic arms. Article V, paragraph 3 of our draft agreement requires parties to locate, inventory, and consolidate covered chip clusters within 120 days.

Consolidating assets to facilitate verification of compliance is often another step in restrictive treaties. \href{https://1997-2001.state.gov/www/global/arms/starthtm/start/abatext.html#art3}{Article III} of START I forbade ICBMs from being co-located with space-launch facilities, easing monitoring. Paragraph 1, subparagraph (a) commits parties to “avoid co-locating AI chips with non-ancillary non-AI computer hardware in these declared facilities” for the same reason.

Another reason to pursue consolidation is to limit breakout potential (breakout is when a party attempts to create ASI before anyone else can stop them). Consolidated assets are easier to surveil and disable if necessary. The anticipation of a credible disabling action improves the effectiveness of deterrence-by-denial.

Monitoring and inspections are common components of prior treaties in limited-trust contexts; we have consequently drafted provisions for this where appropriate, in paragraphs 1, 4, 6, and 7 of this article. Some specific precedent for this:
\begin{itemize}[nosep]
    \item Verification of START I included \href{https://www.armscontrol.org/factsheets/start-i-glance}{hundreds of on-site inspections} in the first few years.
    \item The CWC requires the declaration and inspection of all Chemical Weapons Production Facilities — there have been 97 \href{https://www.opcw.org/media-centre/opcw-numbers}{declared} — and the majority of these have been verifiably destroyed. (In requiring the declaration of existing facilities, these agreements also prohibit certain activities from occurring outside declared facilities, analogous to this article’s prohibition on unmonitored CCCs.)
    \item Over \href{https://www.armscontrol.org/factsheets/iaea-safeguards-agreements-glance}{700 declared nuclear facilities} around the world are monitored by the IAEA as part of the NPT.
\end{itemize}

Similar to paragraph 3 of this article, numerous arms control agreements require that parties not interfere with each other’s NTM in the context of verification. Examples include SALT I,\footnote{The Strategic Arms Limitation Talks (SALT) commenced in 1969 between the U.S. and USSR, producing the SALT I treaty, signed in 1972, which froze the number of strategic ballistic missile launchers and regulated the addition of new submarine-launched ballistic missiles, among other restrictions.} ABM,\footnote{The 1972 \href{https://en.wikisource.org/wiki/Anti-Ballistic_Missile_Treaty}{Anti-Ballistic Missile Treaty} (ABM) grew out of the original SALT talks, and limited each party to two anti-ballistic complexes each (later, just one) with restrictions on their armament and tracking capabilities.} INF,\footnote{With the 1987 \href{https://2009-2017.state.gov/t/avc/trty/102360.htm}{Intermediate-Range Nuclear Forces Treaty} (INF), the U.S. and USSR agreed to ban most nuclear delivery systems with ranges in between those of battlefield and intercontinental systems. (Given the short warning time strikes from such systems would afford, they were seen more as destabilizing offensive systems than as defensive assets.)} and START I.

Precedent for parties restricting their domestic private sector industries to meet commitments (as would need to be the case with AI) can be seen in U.S. legislation following its ratification of the CWC: The \href{https://www.congress.gov/105/plaws/publ277/PLAW-105publ277.pdf#page=857}{Chemical Weapons Convention Implementation Act of 1998} and Department of Commerce regulations ensured U.S. entities were in compliance. Similarly, the U.S. Congress \href{https://www.epa.gov/ozone-layer-protection/ozone-protection-under-title-vi-clean-air-act}{amended} the Clean Air Act following ratification of the Montreal Protocol to ban ozone-depleting substances.

Approaches to implementing chip centralization in the U.S. might run through the Fifth Amendment’s \href{https://constitution.congress.gov/browse/essay/amdt5-9-1/ALDE_00013280/}{Takings Clause}, in which the government can use its power of eminent domain to seize private property for public purposes, so long as it pays appropriate compensation.

\paragraph{Notes on Article V}
Article V aims to centralize, into monitored facilities, all AI chip clusters (i.e., sets of interconnected chips above a small size) and the vast majority of AI chips. Monitoring and prevention of proliferation are covered in Article VI.

Our draft specifies international verification of this centralization process so that all parties can confirm that all other parties have also centralized their chips. Verification of this type is likely to be straightforward for large AI data centers, as intelligence agencies are likely to already know where these are. For smaller data centers, the CTB can coordinate oversight of domestic centralization efforts.

Chip centralization is an important first step to restricting the development of artificial superintelligence. Centralizing chips in declared facilities enables further monitoring for how these chips are being used, or verification that they are powered off (if they are not safe to use). Centralization would also make it easier for parties to disable these chips, as might become necessary under Article XII, if a Party persists in violating the agreement.

We avoid recommending, in the text of the agreement, that CCCs be located away from population centers, despite their capacity for danger. We avoid this restriction both because (in the case of agreement violations) data centers can likely be shut down without much collateral damage, and because modern data centers are already regularly located near cities. That said, alternative agreements might prefer to treat AI data centers as military facilities, given their potential to pose grave security threats.

\subparagraph{Verifying Centralization}
Most parties would not and should not blindly trust other parties to follow the rules, and would need some way to verify compliance. The centralization of AI chips into declared facilities makes it possible for inspections and monitoring to confirm the presence and activity of the chips.

Centralization might not be strictly necessary if there are other ways to monitor AI chips. Unfortunately, we think this is currently the only feasible option short of physically destroying all existing stockpiles of AI chips, given the limited security mechanisms in current chips today.

In the future, \href{https://www.rand.org/pubs/working_papers/WRA3056-1.html}{hardware-enabled governance mechanisms} could be developed to enable remote governance of AI chips, so that chips don’t need to be centralized to declared locations. \href{https://www.iaps.ai/research/secure-governable-chips}{Aarne et al. (2024)} provide estimates for the implementation time of some of these on-chip governance mechanisms. Their estimates cover the timeline to develop mechanisms that are robust against different adversaries. For concision, we will use their estimates for security in a covertly adversarial context where competent state actors may try to break the governance mechanisms but would face major consequences if caught. They estimate a development time of two to five years for ideal solutions, with less secure but potentially workable options available in just months.

Even though that report is over a year old, we are not aware of significant progress toward these mechanisms, and we think two to five additional years is the most relevant estimate from Aarne et al. Which is to say that, possibly, after a few years of research and development into chip security measures, it would be possible to confidently monitor chips without centralizing them, after some further lag time for new securely-monitorable chips to be produced, and/or for old chips to be retrofitted. Aarne et al. estimate that the first of these options might take four years, but we are optimistic that retrofitting could be done in one to two years if chips are already being tracked.

While centralization as discussed in Article V entails the physical concentration of covered chip clusters, it does not require that governments take ownership of chips. For large data centers, the agreement permits the data center and its chips to remain where they are, under private ownership, so long as they receive monitoring and oversight from the relevant parties. This monitoring would ensure that data centers are engaged only in non-AI activities or permitted AI activities like running old models. For smaller chip aggregations, it may be necessary to physically move them into a larger data center, with their owner continuing to access the chips remotely; we do not consider this an overly onerous restriction given that it is already common in cloud computing arrangements.

\subparagraph{Feasibility}
It looks feasible to verifiably consolidate the majority of AI chips. The very largest AI data centers, such as those with more than 100,000 H100-equivalents, are hard to hide. They are detectable from their physical footprint and power draw, and many of them are publicly reported on. In fact, it’s probably possible for intelligence services to track and locate data centers as small as around 10,000 H100-equivalents. Locating smaller data centers would involve domestic authorities using various powers in cooperation with CTB inspectors.

States have a range of tools available for tracking down chips owned domestically. They can legally require reporting of all chip clusters larger than 16 H100s; they can use sales records and other financial information from chip distributors; they can  interview technicians with expertise in data center construction; etc. If they suspect smuggling, obfuscation, or concealment of chips, they can employ law enforcement to investigate further. This process of domestic centralization would be overseen by CTB inspectors to ensure thorough compliance.

Locating large data centers could happen quickly, in days or weeks. Actually centralizing chips could take longer, as it might be necessary to build further data center capacity in the facilities that would become CCCs.

One significant challenge is providing justified confidence that one Party is not doing a secret AI project with non-declared AI chips. Chip centralization provides some assurance, but it may not be sufficient if some country could purposefully undermine its own domestic centralization efforts. For further assurances against illegal AI projects, see the intelligence gathering and challenge inspections discussed in Article X.

For a longer discussion of chip consolidation, see Appendix D.

\subparagraph{On the Definition of CCCs}
Our definition of CCC draws a line at 16 H100-equivalents. This threshold aims to meet a few criteria:
\begin{itemize}[nosep]
    \item Monitoring chip clusters larger than 16 H100s works well with the training FLOP thresholds in Article IV. Training with 16 H100s (FP8 precision, 50 percent utilization — realistic but optimistic parameters) would take 7.3 days to get to $10^{22}$ FLOP (the Monitored Threshold), and 2 years to get to $10^{24}$ FLOP (the Strict Threshold). Therefore, it would be feasible for people to use undeclared chips to reach the bottom threshold, but it would be somewhat impractical for them to get to the prohibited training threshold.
    \item This threshold is plausibly sufficient for preventing the advancement of AI capabilities, when combined with bans on AI research in Article VIII. Article IV lays out training restrictions where large-scale training is prohibited and medium-scale training is allowed but subject to oversight. It is probably acceptable — that is, it probably poses minimal risk — to allow small-scale training, such as the amount that can be done on 16 H100s in a realistic time frame.
    \item This threshold has limited impact on hobbyists and consumers. Very few individuals own more than 16 H100s. In mid-2025, a set of 16 H100 chips cost around \$500,000. This isn’t a threshold one would accidentally cross by having a few old gaming consoles laying around.
    \item Consolidating AI chips gets harder as the allowable quantity shrinks. Finding data centers with 100,000 chips is easy; finding those with 10,000 is likely also relatively easy; with 1,000 it’s unclear; and below 100, it may start to become quite difficult. The 16 H100 threshold is likely to be challenging, and is picked partially due to the increasing infeasibility of still lower thresholds.
    \item Despite potential enforcement challenges, it is possible that this definition would need to be revised and the threshold brought lower (e.g., 8 H100-equivalents). In our agreement, the CTB would be tasked with assessing this definition and changing it as needed.
\end{itemize}

\subparagraph{Other Considerations}
This article calls for parties to avoid co-locating AI chips with non-ancillary non-AI chips. This is suggested because co-location might make verification of chip use (Article VII) more difficult. However, this is not strictly necessary, and it may not be desired. AI chips are currently often colocated with non-AI chips, and the inconvenience of changing this could outweigh the inconvenience of monitoring and verifying the AI chips in a data center that mixes AI chips with non-AI chips.

There is some risk that private citizens could construct an unmonitored CCC from “loose” H100-equivalent chips. To combat this, the agreement holds that parties shall make “reasonable effort” to monitor chip sales (in excess of 1 H100-equivalent) and detect the formation of new CCCs. More stringent measures could be taken, such as requiring all such chips and sales to be formally registered and tracked. Our draft does not go to that length, both because we do not expect all that many “loose” H100-equivalent chips to be unaccounted-for after all chips in CCCs are cataloged, and because other mechanisms (such as the whistleblower protections in Article X) help with the detection of newly-formed CCCs.

Rather than immediately requiring small clusters (e.g., 100 H100s) to be centralized, the agreement could instead implement a staged approach. For example: In the first 10 days all data centers with more than 100,000 H100-equivalent chips must be centralized and declared, then in the next 30 days all data centers with more than 10,000 H100-equivalent chips must be centralized and declared, etc. A tiered approach might better track international verification capacity as intelligence services ramp up their detection efforts.

One downside of a staged approach is that it might provide more opportunities for states to hide chips and establish secret data centers. This approach nevertheless parallels how some previous international agreements have worked within the constraints of their verification and enforcement options. For instance, the 1963 Partial Test Ban Treaty did not ban underground testing of nuclear weapons, due to the difficulty in detecting such tests.

\subsection*{ARTICLE VI — AI Chip Production Monitoring}
\begin{enumerate}
    \item The CTB will coordinate monitoring of AI chip production facilities and key inputs to chip production. This monitoring will ensure that all newly produced AI chips are immediately tracked and monitored until they are installed in declared CCCs and that unmonitored supply chains are not established.
    \begin{enumerate}
        \item The CTB will coordinate monitoring of AI chip production facilities determined to be producing or potentially producing AI chips and relevant hardware [the precise definitions of AI chip production facilities, AI chips, and relevant hardware would need to be further described in an Annex; the monitoring methods would also need to be described in an Annex].
        \item Monitoring of newly produced AI chips will include monitoring of production, sale, transfer, and installation. Monitoring of chip production will start with fabrication. The full set of activities includes fabrication of high-bandwidth memory (HBM), fabrication of logic chips, testing, packaging, and assembly [this set of activities would need to be specified in an Annex].
    \end{enumerate}
    \item For facilities where tracking and monitoring is not feasible or implemented, production of AI chips will be halted. Production of AI chips may continue when the CTB declares that acceptable tracking and monitoring measures have been implemented.
    \item If a monitored chip production facility is decommissioned or repurposed, the CTB will coordinate oversight of that process, and, if done satisfactorily, this ends the monitoring requirement.
    \item No Party sells or transfers AI chips or AI chip manufacturing equipment except as authorized and tracked by the CTB.
    \begin{enumerate}
        \item Sale or transfer of AI chips within or between Parties to the Agreement has a presumption of approval and is tracked by the CTB.
        \item Sale or transfer of AI chip manufacturing equipment within or between Parties to the Agreement does not have a presumption of approval. Approval for such transfer requires consensus of the Executive Council, based on an assessment of the risk of diversion or withdrawal from the Agreement of the receiving Party.
        \item Sale or transfer of AI chips and AI chip manufacturing equipment to non-Party States or entities outside a Party State has a presumption of denial.
    \end{enumerate}
    \item No Party sells or transfers non-AI advanced computer chips or non-AI advanced computer chip manufacturing equipment to non‑Party States or entities outside a Party State except as authorized and tracked by the CTB.
    \item Sale or transfer of non-AI advanced computer chips or non-AI advanced computer chip manufacturing equipment within or between Parties to the Agreement is not restricted under this Article.
    \item To prevent accumulation of excess chip production capacity that could enable rapid breakout from the Agreement, the Executive Council may impose limits on total annual production of AI chips. Such limits aim to allow replacement of aging chips and modest expansion for approved applications while preventing stockpiling that would reduce the time required for a Party to develop ASI after withdrawal.
\end{enumerate}

\paragraph{Precedent for Article VI}
Treaty provisions for monitoring production facilities are not new. Article XI of the 1987 INF allowed for thirteen years of inspections of designated facilities where intermediate-range nuclear delivery systems had previously been produced; Section VII of the accompanying \href{https://2009-2017.state.gov/t/avc/trty/102360.htm\#inspections}{inspection protocol} permitted continuous perimeter and portal monitoring that could include weighing (and in some cases x-raying) any vehicle leaving the facility large enough to carry a relevant missile.

Monitoring AI chip production is more complicated, due to the difficulty of discerning a chip’s function and capabilities from outward characteristics; this is why our Article VI stipulates that “relevant hardware would need to be further described in an Annex,” along with monitoring methods. But the experience of IAEA safeguards under the NPT shows that verification of a wide variety of production components and precursors across a supply chain is possible. One way the IAEA does this is by \href{https://www-pub.iaea.org/MTCD/Publications/PDF/Pub1669_web.pdf}{providing guidelines} for the design of facilities to make them inspection friendly and reduce compliance costs.

Transfer embargoes on end-products, precursors, and production equipment (like the one suggested here on sale or transfer of AI chips and advanced computer chip manufacturing equipment to non-Party states or entities) all have substantial precedent:
\begin{itemize}[nosep]
    \item In Article I of the \href{https://www.un.org/en/conf/npt/2005/npttreaty.html}{NPT}, each nuclear-weapon state commits “not to transfer to any recipient whatsoever nuclear weapons or other nuclear explosive devices” In its Article III, paragraph 2, they also agree not to provide a “source or special fissionable material” or equipment “especially designed or prepared for the processing, use or production of special fissionable material.”
    \item Article I of the \href{https://2009-2017.state.gov/t/avc/trty/127917.htm}{CWC} likewise commits parties to never “transfer, directly or indirectly, chemical weapons to anyone”; its Article VII requires them to subject listed precursors to specified “prohibitions on production, acquisition, retention, transfer, and use”
    \item The Cold-War-era \href{https://www.govinfo.gov/content/pkg/GPO-CRPT-105hrpt851/html/ch9bod.html\#anchor5563742}{Coordinating Committee for Multilateral Export Controls} (CoCom) established a coordinated set of export controls from Western Bloc countries to the Communist Bloc, covering nuclear-related materials, munitions, and dual-use industrial items such as semiconductors.
    \item The \href{https://www.nuclearsuppliersgroup.org/index.php/en/}{Nuclear Suppliers Group} is a multilateral export control regime that restricts the supply of nuclear and nuclear-related technology that could be diverted to nuclear weapons programs.
    \item Especially relevant is the series of U.S. \href{https://www.bis.gov/press-release/commerce-strengthens-export-controls-restrict-chinas-capability-produce-advanced-semiconductors-military}{export controls} that have focused on AI chips and advanced chip manufacturing equipment, covering dozens of countries in the last couple years.
\end{itemize}

\paragraph{Notes on Article VI}
The \href{https://cset.georgetown.edu/publication/securing-semiconductor-supply-chains/}{AI chip supply chain} is narrow and specialized, making it feasible to monitor production. The \href{https://www.datacenterdynamics.com/en/news/nvidia-gpu-shipments-totaled-376m-in-2023-equating-to-a-98-market-share-report/}{vast majority} of AI chips are designed by NVIDIA. The most advanced logic chips (the main processor) used in AI chips are almost all fabricated by TSMC — accounting for around \href{https://www.governance.ai/analysis/computing-power-and-the-governance-of-ai}{90 percent} of market share. \href{https://epoch.ai/data/machine-learning-hardware?view=table}{Most AI chips} are fabricated on versions of TSMC’s five-nanometer process node, a node likely only supported by \href{https://www.blackridgeresearch.com/blog/list-of-tsmc-fabs-in-taiwan-arizona-kumamoto}{two or three manufacturing plants}. EUV lithography machines, a critical component in advanced logic chip fabrication, are made \href{https://www.governance.ai/analysis/computing-power-and-the-governance-of-ai}{exclusively} by ASML. High-bandwidth memory (HBM), another key component to AI chips, is dominated by \href{https://www.trendforce.com/news/2024/04/24/news-amid-foundry-overcapacity-competition-for-hbm-intensifies-rapidly/}{two or three} companies. This narrow and technical supply chain would be relatively easy to monitor and hard to clandestinely replicate. We don’t want to overstate things too much—for example, China has an emerging domestic supply chain that produces some notable AI chips—but even with various caveats like this, monitoring existing chip production seems quite feasible.

Monitoring AI chip production would have relatively small spillover effects. While some of the same processes also produce other chips (e.g., smartphone chips), the chips themselves are easily differentiated. Chip design would change over time, but as a snapshot, current AI chips would be identifiable via their large high-bandwidth memory (HBM) capacity and specialized matrix-multiply components, among other factors.

When it comes to monitoring the AI chip supply chain, based on existing \href{https://cset.georgetown.edu/publication/securing-semiconductor-supply-chains/}{bottlenecks}, a good start might be to monitor HBM production, logic die fabrication, and subsequent steps (e.g., packaging, testing, server assembly), along with key inputs such as EUV lithography machines.

Our Article states that sales of AI chips within Party states will have a presumption of approval, but does not indicate this presumption for AI chip manufacturing equipment. Chip sales are likely to have a relatively short-term effect on AI development capacity, as the lifecycle of AI chips is typically \href{https://epoch.ai/data-insights/gpu-frontier-lifespan\#:~:text=For\%20these\%20newer\%20chip\%20designs,from\%200.6\%20to\%203.7\%20years.}{only a few years}. By contrast, chip manufacturing capacity could lead to significant chip production for many years to come, and it would be especially concerning if a country became a Party, built up an AI chip supply chain, and then withdrew from the agreement. Therefore, we suggest more conservative restrictions on chip manufacturing equipment than on chips themselves.

Paragraphs 4 and 5 of this article permit the sale of AI chips and chip manufacturing equipment to Parties but not to non-Party states or other entities. That is, parties accept risks from chip manufacturing and concentration, but only in cases where the chips are subject to monitoring. The ability to manufacture and possess chips without a protective response from other states thus emerges as a positive incentive to join the agreement.

On its own, this does not prevent non-parties from accessing AI chips in Parties remotely (i.e., cloud computing, or \href{https://www.governance.ai/research-paper/governing-through-the-cloud}{Infrastructure-as-a-Service}), but such chips would be under CTB monitoring to ensure they are not being used in violation of Article IV.

Restrictions on non-parties could go further, if need be. For example, non-parties could be banned from remote access to AI chips (i.e., from renting AI chips in Parties via the cloud) or from accessing AI models via APIs.

If monitoring chip production and preventing smuggling were infeasible, another approach would be to ban all production of new AI chips. This approach would run less of a risk of chips being diverted, but it has the cost of losing the value that these chips could have produced in non-research, non-development AI applications. It would still rely on some monitoring of chip production facilities — e.g., to ensure they are only producing non-AI chips or that they are decommissioned. Our agreement design invites chip production to continue due to the large benefits their use might bring, with the dangers mitigated by monitoring (Article VII).

\subsection*{ARTICLE VII — Chip Use Verification}
\begin{enumerate}
    \item Parties accept continuous on‑site verification of total chip usage at declared CCCs. The methods used for verification will be determined and updated by the CTB, in accordance with the process described in Article III. These methods may include, but are not limited to:
    \begin{enumerate}
        \item In-person inspectors
        \item Tamper-proof cameras
        \item Measurements of power, thermal, and networking characteristics (i.e., off-chip monitoring mechanisms)
        \item On-chip hardware-enabled mechanisms, including retrofitted mechanisms and remote deactivation capabilities that allow the Executive Council to disable chips if compliance cannot be verified
        \item Declaration of the workloads and operations of chips by the CCC operator
        \item Rerunning of declared workloads, coordinated by the CTB, to confirm fidelity of declarations
    \end{enumerate}
    \item The aim of this verification is to ensure chips are not being used for prohibited activities, such as large-scale AI training described in Article IV.
    \item In cases where the CTB assesses that current verification methods cannot provide sufficient assurance that the AI hardware is not being used for prohibited activities, AI hardware must be powered off, and its non-operation continually verified by in-person inspectors or other CTB-approved verification mechanisms.
    \item The CTB may impose various restrictions on how chips can operate in order to ensure proper verification. These restrictions may include but are not limited to:
    \begin{enumerate}
        \item Restrictions on the bandwidth and latency between different chips, or between chips and their data center network, in order to distinguish permitted inference from prohibited training.
        \item Restrictions on the number or rate of FLOP/s or memory bandwidth at which chips can operate, in order to distinguish permitted inference from prohibited training or other prohibited workloads.
        \item Restrictions on the numerical precision of chip operations, in order to differentiate AI from non-AI workloads.
    \end{enumerate}
    \item The CTB will coordinate differentiated verification approaches for different CCCs based on their likelihood of being used for AI activities and their sensitivity as relevant to national security.
    \begin{enumerate}
        \item More sensitive facilities might have more technical/automated verification methods, less extensive physical access for foreign inspectors, and enhanced security protocols for inspector access.
    \end{enumerate}
    \item The CTB will lead research and engineering to develop better technologies for chip use monitoring and verification. Parties will support these efforts [more details would be provided in an Annex].
\end{enumerate}

\paragraph{Precedent for Article VII}
In our discussion of precedent for Article VI, we described the continuous monitoring of former intermediate-range missile production sites under the INF treaty, which, while allowing for weighing and non-destructive scanning of vehicles leaving the facilities, did not allow inspectors inside the trucks or the sites themselves. Analogous perimeter monitoring of data centers can provide some clues about operations from power draw, thermal emissions, and network bandwidth. But reasonable assurance that restricted AI operations are not occurring would likely require some combination of the elements we listed under paragraph 1, which includes tamper-proof cameras, on-chip hardware-enabled mechanisms, and in-person inspectors.

Such practices are already routine for the International Atomic Energy Agency, which is \href{https://www.iaea.org/newscenter/news/surveying-safeguarded-material-24/7}{increasingly using around-the-clock surveillance technologies} to supplement inspections:

\begin{displayquote}
Over a million pieces of encrypted safeguards data are collected by over 1400 surveillance cameras, and 400 radiation and other sensors around the world. More than 23 000 seals installed at nuclear facilities ensure containment of material and equipment.
\end{displayquote}

One of the methods used under START I to verify compliance with missile performance characteristics was the sharing of almost all telemetry data transmitted from in-flight sensors during tests, as specified in the telemetry protocol, which also required parties to provide any playback equipment and data formatting information necessary to interpret it. Depending on the mix of verification methods adopted, some Parties may use analogous methods, building on the light-touch monitoring that is \href{https://www.governance.ai/research-paper/governing-through-the-cloud}{common practice} for cloud computing providers to collect about customer workloads.

Continuous government monitoring of private commercial facilities (as most data centers are) also has plenty of precedent. The U.S. Nuclear Regulatory Commission, tasked with overseeing domestic nuclear reactor safety, places \href{https://www.nrc.gov/reactors/operating/oversight/rop-description/resident-insp-program.html}{two resident inspectors} in each U.S. commercial power plant, and U.S. meat producers \href{https://www.fsis.usda.gov/sites/default/files/media_file/2021-02/Fed-Food-Inspect-Requirements.pdf}{cannot conduct slaughter operations} if inspection personnel from the FSIS\footnote{The Food Safety and Inspection Service (FSIS) is an agency of the U.S. Department of Agriculture formed in 1977.} are not on site to oversee them.

\paragraph{Notes on Article VII}
Parties would want to ensure that existing AI chips are not being used to do dangerous AI training. There are legitimate reasons to use these chips to run existing AI services like (extant versions of) ChatGPT. The agreement thus requires the ability to verify that AI chips are only being used for permitted activities.

This article creates a positive incentive to join the agreement: A country may continue using AI chips as long as supervision can verify that their use does not put the world at risk. Given the goal of preventing large-scale AI training, there are two main approaches: Ensure nobody has the necessary hardware (i.e., that AI chips do not exist), or ensure that the hardware is not used in the development of superintelligence (i.e., via monitoring). Monitoring is what permits the continued safe use of AI chips. This is conceptually analogous to IAEA Safeguards: In order for a non-nuclear weapon country to be permitted nuclear materials and facilities, it is necessary for the IAEA to inspect and ensure the use is only for peaceful purposes.

\subparagraph{Feasibility}
Various technical methods could be used to make verification easier. For example, using the algorithms of 2025, AI training requires much higher bandwidth compared to AI inference. Thus, if the chips are connected using low-bandwidth networking cables, they are effectively limited such that they can engage in inference but not training. There are various nuances to these and other mechanisms; we refer curious readers to \href{https://techgov.intelligence.org/research/mechanisms-to-verify-international-agreements-about-ai-development}{previous work} on \href{https://www.rand.org/pubs/working_papers/WRA4077-1.html}{the topic}.

This article tasks the CTB with developing and implementing better verification mechanisms, defined broadly. We think this flexibility is necessary due to the pace of change in AI and the possibility that unanticipated developments could disrupt verification methods. The state of AI verification research is also nascent; more technological development in verification technology is a key opportunity.

It is much easier to verify whether a new AI is being created than it is to verify that an existing AI is not performing dangerous inference tasks (such as research that advances the creation of superintelligence). As of August 2025, existing AIs don’t obviously seem capable enough for their inference activities to substantially advance the creation of superintelligence, and so the monitoring challenge is easier.

It is unclear how difficult it would be to monitor AI inference activities. Inference monitoring is already applied by many AI companies today, for instance to detect if users are trying to use AIs to make \href{https://openai.com/index/preparing-for-future-ai-capabilities-in-biology/}{biological weapons}, but it is unclear whether that monitoring is comprehensive, and it is unclear whether it would get less reliable if AIs were allowed to become more capable. The longer that AI capabilities are allowed to advance before an agreement resembling our draft comes into effect, the more difficult monitoring would become. Verification that chips are only being used for permitted purposes would become more difficult and more expensive, or might even become impossible.

\subparagraph{Other Considerations}
In theory, verification could be facilitated by technological means that allow for remote monitoring. However, current technology likely contains security vulnerabilities that would allow chip owners to bypass monitoring measures. Thus, verification would likely require either continuous on-site monitoring or that chips be shut off until the technological means mature. Once monitoring technology is mature, strong hardware-enabled governance mechanisms could allow chips to be monitored remotely with confidence.\footnote{Another key consideration for chip use verification measures is security and privacy. Parties will want to ensure that the CTB only has access to the information it needs for verification without also having access to sensitive data on the chips (such as military secrets or sensitive user data). Therefore, the verification methods used would need to be made secure and would be narrowly scoped when possible.}

Paragraph 5 of this article allows for different verification methods for different CCCs. One reason for this discrimination is practical: Different CCCs would require different verification approaches in order to establish justified confidence that they are not being used for dangerous AI development. For example, large data centers that were previously being used for frontier AI training would have the greatest ability to contribute to prohibited training and so might require greater monitoring.

Second, discrimination in verification approaches would make the agreement more palatable by requiring less invasive monitoring for sensitive CCCs. For example, intelligence agencies or militaries may not want any monitoring of their data centers (which may have more computing power than 16 H100-equivalents despite being used for purposes that have nothing to do with AI), and this provision helps strike a balance. It would still be necessary to verify that these data centers are not being used for dangerous AI activities, and the Executive Council members would work through the CTB with these groups to ensure it can get the information it needs while also meeting the privacy and security needs of CCC owners. On the other hand, allowing different verification protocols might hurt the viability of the agreement if it is viewed as unfair.

Our draft agreement allows chip use and production to continue so that the world may benefit from such chips. One alternative approach is to shut down new chip production and/or destroy existing chips. Absent algorithmic advancements, the destruction of chips would increase the “breakout time” — the time it takes between when a group starts trying to create a superintelligence and the point at which they succeed. This is because (in lieu of algorithmic advancements), a rogue actor would need to develop the capability to produce chips, which is a lengthy and conspicuous process. However, because we think it’s feasible to track chips and verify their usage, we do not think that the benefit of longer breakout times is clearly worth the cost of shutting off all AI chips.

\subsection*{ARTICLE VIII — Restricted Research: AI Algorithms and Hardware}
\begin{enumerate}
    \item For the purpose of preventing the development of artificial superintelligence, this Agreement restricts only research that would materially advance toward ASI or undermine verification of compliance with this Agreement. This includes research in the field of machine learning and research in other artificial intelligence paradigms. Research focused on specific applications (such as medical diagnosis, scientific discovery, or industrial automation) that does not advance general cognitive capabilities toward ASI levels is not restricted. Restricted research includes:
    \begin{enumerate}
        \item Improvements to methods for training general-purpose AI systems that would significantly increase model capabilities toward superintelligent performance or dramatically reduce the computational resources required to develop such systems
        \item Distributed or decentralized training methods that would enable ASI development outside of monitored facilities, or training methods specifically optimized to evade the computational thresholds in Article IV
        \item Advancements in the fabrication of AI-relevant chips or chip components
        \item Design of more performant or more efficient AI chips
    \end{enumerate}
    \item Application-specific AI research and development that does not advance general cognitive capabilities is permitted and encouraged. This includes research in domains such as medical diagnostics, drug discovery, materials science, climate modeling, robotics for specific tasks, and other specialized applications.
    \item The CTB's Research Controls division classifies all restricted research activities as either controlled or prohibited.
    \begin{enumerate}
        \item Each Party monitors any controlled research activities within its jurisdiction, and takes measures to ensure that all controlled research is monitored and made available to the Research Controls division for review and monitoring purposes.
        \item Each Party does not conduct any prohibited  research, and prohibits and prevents prohibited  research by any entity within its jurisdiction.
    \end{enumerate}
    \item Parties to the Agreement must not assist, encourage, or share prohibited  research, including by funding, procuring, hosting, supervising, teaching, publishing, providing controlled tools or chips, or facilitating collaboration.
    \item Each Party provides a representative to the CTB's Research Controls division (established in Article III). This division has these responsibilities:
    \begin{enumerate}
        \item Interpret and clarify the categories of restricted research, and respond to questions as to the boundaries of restricted research, in response to new information, and in response to requests from researchers or organizations, or Party members.
        \item Interpret and clarify the boundary between controlled research and prohibited  research, and respond to questions as to this boundary, in response to new information, and in response to requests from researchers or organizations or Party members.
        \item Modify the definition of restricted research and its categories, in response to changing conditions, or in response to requests from researchers or organizations or Party members.
        \item Modify the boundary between controlled research and prohibited  research in response to changing conditions, or in response to requests from researchers or organizations or Party members.
        \item The CTB may modify the categories, boundaries, and definitions of restricted research in accordance with the process described in Article III.
    \end{enumerate}
\end{enumerate}

\paragraph{Precedent for Article VIII}
Pre-emptive restrictions on the dissemination of information related to dangerous technology find precedent in the \href{https://www.atomicarchive.com/resources/documents/postwar/atomic-energy-act.html}{Atomic Energy Act of 1946}, still in force, which established information on certain topics as Restricted Data by default (the “born secret” doctrine); exclusions were at the discretion of the new Atomic Energy Commission created by this legislation:\footnote{The 1946 Atomic Energy Act was later augmented by the \href{https://www.govinfo.gov/content/pkg/COMPS-1630/pdf/COMPS-1630.pdf}{Atomic Energy Act of 1954} with the goal of allowing for a civilian nuclear industry, which required allowing some Restricted Data to be shared with private companies.}

\begin{displayquote}
The term “restricted data” as used in this section means all data concerning the manufacture or utilization of atomic weapons, the production of fissionable material, or the use of fissionable material in the production of power, but shall not include any data which the Commission from time to time determines may be published without adversely affecting the common defense and security.
\end{displayquote}

Unlike other types of government classification, Restricted Data can be created (deliberately or accidentally) by the private sector, a matter of unresolved constitutionality\footnote{The 1979 case of \href{https://en.wikipedia.org/wiki/United_States_v._Progressive,_Inc.}{United States v. The Progressive}, in which a newspaper intended to reveal the “secret” of the hydrogen bomb, might have given the U.S. Supreme Court an opportunity to rule on whether the “born secret” doctrine violates the First Amendment’s protections on speech, if the government hadn’t dropped the case as moot.} that highlights the need for a regulatory arm authorized and capable of making everyday decisions about the exact boundaries of Restricted Data. The \href{https://www.usa.gov/agencies/national-nuclear-security-administration}{National Nuclear Security Administration} (NNSA) does this for nuclear secrets in the U.S. Under our Article VIII, paragraph 5, the Research Controls division would take on this role for restricted AI research. It would also fill other NNSA-analogous functions, outlined in our Article IX, by (1) maintaining relationships with researchers and organizations working on projects that approach the classification threshold, and (2) establishing secure infrastructure for reporting and containment of inadvertent discoveries.

There is also precedent for containing and controlling research in dangerous fields. In the final months of World War II, the U.K. and U.S. collaborated on the \href{https://ahf.nuclearmuseum.org/ahf/history/alsos-mission/}{Alsos Mission} to capture German nuclear scientists, gather information about German progress toward an atomic bomb, and prevent the USSR from obtaining these resources for its own nuclear program. \href{https://airandspace.si.edu/stories/editorial/project-paperclip-and-american-rocketry-after-world-war-ii}{Project Overcast} (also called Operation Paperclip) was a secret U.S. program to take German rocket engineers into U.S. employment after the war.

Containment of restricted AI research within Party states might run through existing regulatory frameworks. In the U.S., these include:
\begin{itemize}[nosep]
    \item The “\href{https://www.bis.gov/learn-support/deemed-exports/what-deemed-export}{deemed exports}” concept in export control law, which obliges a U.S. entity to obtain an export license from the Bureau of Industry and Security\footnote{An arm of the U.S. Department of Commerce.} before sharing controlled technologies with foreign persons by deeming such sharing as an export.
    \item The \href{https://www.pmddtc.state.gov/ddtc_public?id=ddtc_kb_article_page\&sys_id=24d528fddbfc930044f9ff621f961987}{International Traffic in Arms Regulations} (ITAR), a set of U.S. State Department regulations that control the export of military and some dual-use technologies. ITAR was used to prevent the broader development and use of cryptographic techniques by the private sector until 1996, as these were classified as a “defense article” on the \href{https://www.ecfr.gov/current/title-22/chapter-I/subchapter-M/part-121}{United States Munitions List}.
    \item The \href{https://www.congress.gov/bill/82nd-congress/house-bill/4687/text}{Invention Secrecy Act of 1951}, which gives U.S. government agencies the power to impose “secrecy orders” on new patent applications with national security implications. Inventors can not only be denied patents, but legally prohibited from disclosing, publishing, or even using their inventions.\footnote{Hundreds of such orders have been placed on cryptography-related patents over the decades.}
\end{itemize}

Project Overcast also provides precedent for controlling researchers by simply paying them well to act in the interest of the state. Additional precedent for such incentives is discussed with Article IX.

The \href{https://www.pnas.org/doi/epdf/10.1073/pnas.72.6.1981}{1975 Asilomar Conference on Recombinant DNA} is evidence that scientists themselves may voluntarily agree to research prohibitions, especially when faced with novel threats they are not sure they can contain. This conference resulted in voluntary guidelines around recombinant DNA research. These guidelines included prohibitions on certain especially dangerous experiments, such as cloning recombinant DNA from highly pathogenic organisms and DNA containing toxin genes.

\paragraph{Notes on Article VIII}
Prohibiting several broad categories of research, when relevant know-how is already distributed in the private sector, presents a challenge. In our draft, research is restricted if it advances AI capabilities or performance, or if it endangers the verification approach laid out in previous articles.

Some research must be prohibited  to prevent AI capabilities from advancing, even when holding the amount of training FLOP constant. This prohibition would need to cover all research that might make AIs more efficient to train or that might increase the capabilities of AIs, often referred to as “algorithmic progress.” In current paradigms, this includes advances in the algorithms used in pre-training, post-training, and inference. The AI development paradigm might shift and other AI development paradigms might catch up to machine learning. Therefore, the agreement should not constrain itself to only research in machine learning, even if this is the most urgent priority today. The agreement keeps the door open to potentially restrict research in other AI paradigms (e.g., connectomics, brain-inspired AI, fast genetic algorithms, GOFAI), if research in those paradigms seems likely to lead to ASI.

Previous algorithmic innovations, such as the development of the transformer architecture, demonstrate the potential for rapid advances in AI capabilities. Continued innovation could dramatically lower the amount of computational resources required for a given level of AI capability. As a feasibility argument, observe that modern AIs are much less data-efficient than human beings, which suggests that much more data-efficient algorithms can be found.

It is much harder to prevent the training of dangerous AIs when they can be trained with a small number of AI chips, or with many chips geographically dispersed in small clusters.

Separately, a prohibition must preclude research into new ways to manufacture untracked AI chips. Monitoring and verification of AI chips is feasible in large part because of the present complexity and centralization of advanced AI-relevant semiconductor manufacture.

Article VIII also bans research into the design of more performant or efficient AI chips, which otherwise become \href{https://epoch.ai/data-insights/ml-hardware-energy-efficiency}{substantially more efficient} year over year. A data center using more efficient AI chips would be easier to conceal, as these chips would use less electricity for the same or greater performance.

The specific types of research that are restricted would need to be updated in response to changing conditions. One example of such an activity is research into consumer hardware that can efficiently perform AI training activities, if such progress would pose a risk to verification.

Domestic efforts to restrict research could start by focusing on the publication and funding of research. Most researchers want to be law-abiding, gainfully employed citizens; steps that push dangerous AI research outside of accepted social norms would likely be impactful.

The diversity of restricted actions in paragraph 4 addresses a need to ensure that if research activities are split between multiple jurisdictions, the agreement still unambiguously holds each state responsible for prohibiting and preventing the individual activities. Paragraph 4 applies, for example, in the case where a company in one jurisdiction hires an employee in a second who remotely operates chips hosted in a third.

\subsection*{ARTICLE IX — Research Restriction Verification}
\begin{enumerate}
    \item Each Party creates or empowers a domestic agency with the following responsibilities:
    \begin{enumerate}
        \item Maintain awareness of and relationships with domestic researchers and organizations working on areas adjacent to restricted research, in order to communicate the categories of restricted research established in Article VIII.
        \item Impose penalties to deter domestic researchers and organizations from conducting restricted research. These penalties are proportionate to the severity of the violation and are designed to act as a sufficient deterrent. Each Party enacts or amends legal statutes as necessary to enable the imposition of these penalties.
        \item Establish secure infrastructure for reporting and containment of inadvertent discoveries meeting the conditions for restricted research. These reports will be shared with the Research Controls division.
    \end{enumerate}
    \item To aid in the international verification of research bans, the Research Controls division will develop and implement verification mechanisms.
    \begin{enumerate}
        \item These mechanisms could include but are not limited to:
        \begin{enumerate}
            \item Interviews of researchers who have previously worked in restricted research topics, or are presently working in adjacent areas, conducted by the U.S. and China and coordinated by the Research Controls division. These interviews may be overseen by the researcher’s home state to ensure no misconduct.
            \item Monitoring of the employment status of researchers who have previously worked in restricted research topics, or are presently working in adjacent areas.
            \item Maintaining embedded auditors provided by the U.S. and China in selected high-risk organizations (e.g., projects difficult to distinguish from restricted research, organizations that were previously AI research organizations).
        \end{enumerate}
        \item Parties, in particular, the U.S. and China, assist in the implementation of these verification mechanisms.
        \item The information gained through these verification mechanisms will be compiled into reports for the Executive Council, keeping as much sensitive information confidential as possible to protect the privacy and secrets of individuals and Parties.
    \end{enumerate}
\end{enumerate}

\paragraph{Precedent for Article IX}
Existing agencies empowered to “maintain awareness of and relationships with domestic researchers and organizations” at risk of developing restricted information, as called for by paragraph 1, subparagraph (a), include the DOE and NNSA, discussed in the precedent section for Article VIII.

Precedent for “monitoring of the employment status of researchers” in high-risk fields, as we suggest in paragraph 2, subparagraph (a)(ii), can be found in the International Science and Technology Center (ISTC).\footnote{The International Science and Technology Center grew out of the 1991 \href{https://en.wikipedia.org/wiki/Nunn\%E2\%80\%93Lugar_Cooperative_Threat_Reduction}{Nunn-Lugar Cooperative Threat Reduction} program, a U.S. initiative to secure and dismantle WMDs and their associated infrastructure in former Soviet states.} Established in 1994, the ISTC was specifically created to reduce nuclear proliferation risks by \href{https://astanatimes.com/2014/12/istc-headquartered-nazarbayev-university-2015/}{keeping Soviet nuclear researchers gainfully employed in peaceful} activities and connected to the international scientific community. The ISTC also shows the potential of incentives as a complement to penalties for keeping technical experts (who may find themselves unemployed as a result of this agreement) from engaging in restricted research.

Penalties may need to be severe to provide the deterrence indicated in our Article IX, paragraph 1, subparagraph (b). While it may not be directly applicable here, there is precedent for severely punishing the sharing of sensitive data in some contexts. The \href{https://www.govinfo.gov/content/pkg/COMPS-1630/pdf/COMPS-1630.pdf\#page=154}{Enforcement chapter (18)} of the 1946 Atomic Energy Act includes fines and imprisonment for unauthorized disclosures. More severe punishments are sometimes levied for other information-disclosure crimes, namely treason.\footnote{Parties to our agreement may wish to explore expanding the concept of \href{https://www.law.cornell.edu/wex/crime_against_humanity}{crimes against humanity} (codified in the \href{https://en.wikisource.org/wiki/Rome_Statute_of_the_International_Criminal_Court}{1998 Rome Statute of the International Criminal Court}) to cases where a researcher deliberately seeks to develop ASI at the expense of the people of Earth.}

When developing secure “infrastructure for reporting and containing inadvertent discoveries of restricted research,” precedent and potentially usable templates may be found in the extensive DOE procedures for handling different categories of sensitive data. The DOE’s Occurrence Reporting and Processing System, as well as the Committee on National Security Systems’s\footnote{The \href{https://en.wikipedia.org/wiki/Committee_on_National_Security_Systems}{Committee on National Security Systems} (CNSS) is a U.S. intergovernmental organization that sets security policies for government information systems.} instructions for classified information spillage, may also be of use.

The Research Controls division might look to existing practices by the IAEA when developing inspection protocols. Under the framework of the Model Additional Protocol approved in 1997 by the IAEA Board of Governors, states that have made comprehensive safeguard agreements\footnote{144 States, as of June 2025} allow complementary access inspections that look for undeclared nuclear material. As part of such visits, inspectors may interview operators, analogous to our proposal in paragraph 2, subparagraph (a)(i).

To “protect the privacy and secrets of individuals and parties” when performing verifications, as required by this article’s paragraph 2, subparagraph (c), the Research Controls division might adapt compartmentalization practices of parties’ existing intelligence agencies and multilateral intelligence-sharing agreements. For example, under the “third party rule” or “originator control principle” understood to be commonplace in such arrangements, it is prohibited to disclose shared information to third parties (potentially even oversight bodies) without permission from the originating agency.

\paragraph{Notes on Article IX}
To help verify that there is no prohibited AI research happening, Article IX tasks parties with  demarcating “areas adjacent to restricted research” and then establishing relationships with the researchers working in these adjacent areas. There are sufficiently few top AI researchers in the world that it may be feasible to track the activities of a significant fraction of them. The technical staff of top AI companies numbers on the order of 5,000 researchers\footnote{OpenAI’s \href{https://openai.com/index/introducing-gpt-5/}{GPT-5 announcement} lists 470 contributors. Google’s \href{https://arxiv.org/pdf/2403.05530\#page=92.10}{Gemini 1.5 technical report} lists 1135 contributors. The \href{https://arxiv.org/abs/2407.21783}{Llama 3 paper} lists 559 authors. The \href{https://arxiv.org/abs/2412.19437v1}{DeepSeek-V3 paper} lists 199 authors. The \href{https://arxiv.org/abs/2505.09388}{Qwen3 paper} lists 177 total contributors. The \href{https://arxiv.org/abs/2507.20534}{Kimi k2 paper} lists 168 authors. The sum of these numbers is 2,708. Other frontier AI companies such as Anthropic and xAI do not provide sufficient details about the number of technical contributors, to our knowledge. There are other companies that may have sufficient AI talent, therefore we conservatively estimate that the total number of researchers at top AI companies is on the order of 5,000.}, and it is commonly believed that a much smaller group is critical to frontier AI development, likely numbering in the hundreds.\footnote{In a \href{https://www.theverge.com/decoder-podcast-with-nilay-patel/761830/amazon-david-luan-agi-lab-adept-ai-interview?utm_source=chatgpt.com}{2025 interview}, David Luan, head of Amazon’s AGI research lab, estimated the number of people he would trust “with a giant dollar amount of compute” to develop a frontier model at “sub-150.”} The number of attendees of top AI conferences is \href{https://ourworldindata.org/grapher/attendance-major-artificial-intelligence-conferences}{estimated} to be about 70,000. As higher-end estimates, the number of employees at relevant hardware companies likely numbers around a million\footnote{\href{https://www.macrotrends.net/stocks/charts/TSM/taiwan-semiconductor-manufacturing/number-of-employees}{TSMC employee count}: 84,000 in 2025. \href{https://www.macrotrends.net/stocks/charts/INTC/intel/number-of-employees}{Intel employee count}: 125,000 in 2023. \href{https://news.samsung.com/global/fast-facts}{Samsung employee count}: 263,000 in 2024. \href{https://www.smics.com/en/site/company_info}{SMIC employee count}: 20,000 in 2023. \href{https://www.macrotrends.net/stocks/charts/NVDA/nvidia/number-of-employees\#:~:text=NVIDIA\%20total\%20employee\%20count\%20in,End\%20of\%20interactive\%20chart.}{NVIDIA employee count}: 36,000 in 2025. \href{https://www.macrotrends.net/stocks/charts/AMD/amd/number-of-employees}{AMD employee count}: 26,000 in 2023. \href{https://www.huawei.com/en/media-center/company-facts}{Huawei employee count}: 208,000 in 2024. Google TPU division: unknown. \href{https://www.asml.com/en/company/about-asml}{ASML employee count}: 44,000 in 2024. \href{https://news.skhynix.com/corporate/fact-sheet/}{SK Hynix employee count}: 47,000 in 2024. \href{https://www.macrotrends.net/stocks/charts/MU/micron-technology/number-of-employees}{Micron employee count}: 43,000 in 2023. The sum of these numbers is 896,000. So the total number of employees at the most relevant AI hardware companies is likely around one million, though the core technical staff are probably a small subset of this and the number of crucial researchers may only be a few thousand.} and the number of people with at least basic technical AI knowledge is likely in the single-digit or tens of millions\footnote{In April 2025, the popular platform for sharing AI models and datasets \href{https://huggingface.co/spaces/huggingface/how-to-upgrade-to-enterprise}{HuggingFace had} 8 million users. The number of students who have taken \href{https://www.coursera.org/specializations/machine-learning-introduction}{Andrew Ng’s Machine Learning course} is about 4.8 million. The number of developers using GitHub (broader than just AI) was \href{https://github.blog/news-insights/company-news/100-million-developers-and-counting/}{around} 100 million in 2023, and the total number of professional software developers in the world is \href{https://www.slashdata.co/post/global-developer-population-trends-2025-how-many-developers-are-there}{estimated} at 47 million in 2025.}. States could interview researchers about their activities and offer asylum and financial incentives for any whistleblowers (see Article X).

While much about current AI development practices happens in the public view, we think legal restrictions would dramatically hamper the efforts of rogue actors to create superintelligent machines.

Monitoring could be extended to researchers and engineers involved in semiconductor design and manufacture if states are willing to incur the extra costs. A more affordable alternative might be to monitor semiconductor manufacturing companies rather than individuals, taking advantage of complex dependencies within the industry which ensure that small groups of rogue individuals would have trouble creating their own chip fabricators.

Parties may be concerned that other parties will violate domestic research bans and hide research efforts from foreign intelligence. Most likely, large efforts involving many researchers and AI-relevant chips would be noticed by a determined intelligence community. But smaller efforts, like developing alternative machine intelligence paradigms, might only involve a few researchers and commonly available hardware. Verifying research controls is a complex and sensitive undertaking requiring ongoing effort and iteration. To facilitate that end, Article X (below) institutes a variety of tools to facilitate intelligence gathering and to protect whistleblowers.

\subsection*{ARTICLE X — Information Consolidation and Challenge Inspections}
\begin{enumerate}
    \item A key source of information for the coalition is the independent information gathering efforts of Parties. As such, the Information Consolidation division (Article III) will be ready to receive this information. This division coordinates verification and monitoring activities conducted by Parties. Parties conduct monitoring, inspections, and verification using their own capabilities, including intelligence community resources. The CTB establishes standards and protocols for these activities and serves as the central point for receiving declarations and sharing information.
    \begin{enumerate}
        \item The Information Consolidation division takes precautions to protect commercial, industrial, security, and state secrets and other confidential information coming to its knowledge in the implementation of the Agreement, including the maintenance of secure, confidential, and, optionally anonymous reporting channels.
        \item For the purpose of providing assurance of compliance with the provisions of this Agreement, each Party uses National Technical Means (NTM) of verification at its disposal in a manner consistent with generally recognized principles of international law.
        \begin{enumerate}
            \item Each Party undertakes not to interfere with the National Technical Means of verification of other Parties operating in accordance with the above.
            \item Each Party undertakes not to use deliberate concealment measures which impede verification by national technical means of compliance with the provisions of this Agreement.
            \item Parties are encouraged, but not obligated, to cooperate in the effort to detect dangerous AI activities in non-Party countries. Parties are encouraged, but not obligated, to support the NTM of Parties directed at non-Parties, as relevant to this Agreement.
        \end{enumerate}
    \end{enumerate}
    \item A key source of information for the coalition is individuals who provide evidence of dangerous AI activities to the coalition. These individuals are subject to whistleblower protections.
    \begin{enumerate}
        \item This Article establishes protections, incentives, and assistance for individuals ("Covered Whistleblowers") who, in good faith, provide the coalition or a Party with credible information concerning actual, attempted, or planned violations of this Agreement or other activities that pose a serious risk of human extinction, including concealed chips, undeclared data centers, prohibited training or research, evasion of verification, or falsification of declarations. Covered Whistleblowers include employees, contractors, public officials, suppliers, researchers, and other persons with material information, as well as Associated Persons (family members and close associates) who assist or are at risk due to the disclosure.
        \item Parties prohibit and prevent retaliation against Covered Whistleblowers and Associated Persons, including but not limited to dismissal, demotion, blacklisting, loss of benefits, harassment, intimidation, threats, civil or criminal actions, visa cancellation, physical violence, imprisonment, restriction of movement, or other adverse measures. Any contractual terms (including non‑disclosure or non‑disparagement agreements) purporting to limit protected disclosures under this Agreement are void and unenforceable. Mistreatment of whistleblowers constitutes a violation of this Agreement and is handled under Article XI, paragraph 3.
        \item The CTB maintains secure, confidential, and, optionally anonymous reporting channels. Parties establish domestic channels interoperable with the CTB system. The CTB and Parties protect the identity of Covered Whistleblowers and Associated Persons and disclose it only when strictly necessary and with protective measures in place. Unauthorized disclosure of protected identities constitutes a violation of this Agreement and is handled under Article XI, paragraph 3.
        \item Parties offer asylum or humanitarian protection to Covered Whistleblowers and their families, provide safe‑conduct travel documents, and coordinate secure transit.
    \end{enumerate}
    \item Challenge inspections of suspected sites may be conducted upon credible information about dangerous AI activities.
    \begin{enumerate}
        \item Parties may request a challenge inspection. The Executive Council, either by request or because of the analysis provided by the Information Consolidation division, will consider the information at hand in order to request additional information, of Parties or non-Parties, or to propose a challenge inspection, or to decide that no further action is warranted.
        \item A challenge inspection requires approval by consensus of the Executive Council.
        \item Access to a suspected site must be granted by the nation in which the site is present within 24 hours of the Executive Council calling for a challenge inspection. During this time, the site may be surveilled, and any people or vehicles leaving the site may be inspected by officials from a Party.
        \item The challenge inspection will be conducted by a team of inspectors who are approved by both the Party being inspected and the Party that requested the inspection. The CTB is responsible for working with Parties to maintain lists of approved inspectors for this purpose.
        \item Challenge inspections may be conducted in a given Party's territory at most 20 times per year, and this limit can be changed by consensus of the Executive Council.
        \item Inspectors will take absolute care to protect the sensitive information of the inspected Party, passing along to the Executive Council only what information is pertinent to the Agreement.
    \end{enumerate}
    \item U.S. and PRC Independent Verification Efforts
    \begin{enumerate}
        \item The U.S. and PRC each operate an independent verification effort to assess compliance with this Agreement. These verification efforts build upon the pre-existing capabilities of each member's intelligence community and supplement the work of the Coalition Technical Body.
        \item The Coalition Technical Body provides all information it receives or generates to the U.S. and the PRC to support their independent verification efforts. This includes but is not limited to: declarations, reports, monitoring data, inspection results, registers, and any other information relevant to compliance with this Agreement. The CTB does not filter or withhold information from the U.S. and PRC.
    \end{enumerate}
\end{enumerate}

\paragraph{Precedent for Article X}
We previously discussed precedent for information consolidation with Article VIII, where we cited the existence of intelligence agreements understood to include compartmentalization practices like the “third party rule.” Similar rules can be seen in the IAEA, as in \href{https://www.iaea.org/sites/default/files/publications/documents/infcircs/1972/infcirc153.pdf}{INFCIRC/153} Part 1.5:

\begin{displayquote}
…the Agency shall take every precaution to protect commercial and industrial secrets and other confidential information coming to its knowledge in the implementation of the Agreement.
\end{displayquote}

Staff are bound by confidentiality obligations and face criminal penalties for leaks. This matters, because the IAEA has benefited from the intelligence disclosures of participating states, including satellite imagery and documents, \href{https://carnegieendowment.org/posts/2015/12/iran-and-the-evolution-of-safeguards?lang=en}{as in the case of Iran’s} undeclared enrichment activities. Similarly, the IAEA required a special inspection of North Korea’s undeclared plutonium production in \href{https://www.nonproliferation.org/wp-content/uploads/npr/dembin22.pdf\#page=4}{response to provided intelligence}.

Recognizing the indispensable role of national technical means (NTM — satellite imagery, signals collection, and other remote sensing) in verification of multilateral agreements, our draft agreement borrows language from the ABM treaty limiting anti-ballistic missile systems, in which “each Party shall use national technical means of verification” and “undertakes to not interfere with the national technical means of verification of the other Party.” Similar language can be found in Article XII of the 1987 \href{https://2009-2017.state.gov/t/avc/trty/102360.htm}{Intermediate-Range Nuclear Forces Treaty}, Article IV of the 1996 \href{https://www.ctbto.org/sites/default/files/2023-10/2022_treaty_booklet_E.pdf\#page=28}{Comprehensive Nuclear-Test-Ban Treaty}, and throughout the 2010 \href{https://2009-2017.state.gov/documents/organization/140047.pdf}{New START treaty}.

As NTM would not be sufficient for detecting all dangerous violations in the case of ASI, we have borrowed features of the \href{https://www.iaea.org/topics/safeguards-legal-framework}{IAEA Safeguards framework} that encourage internal reporting and provide channels for doing so. But these are hampered by a lack of explicit whistleblower protections; nothing in the NPT or these Safeguards would protect an informant from their government if it decides to retaliate unless that state has applicable domestic protections in place. The provisions for whistleblower protection and asylum in our draft agreement are meant to address this shortcoming.

Recent EU legislation on AI has taken similar measures. The EU AI Act’s \href{https://artificialintelligenceact.eu/recital/172/}{Recital 172} explicitly extends the Union’s existing general \href{https://eur-lex.europa.eu/eli/dir/2019/1937/oj/eng}{whistleblower protections} to those reporting AI Act infringements.

The \href{https://www.ohchr.org/en/instruments-mechanisms/instruments/convention-relating-status-refugees}{1951 Refugee Convention} provides a possible framework for granting asylum to informants, basing qualification on “well-founded fear of being persecuted,” though an amendment or supplemental agreement may be needed to ensure that AI whistleblowing is a legally qualifying cause of persecution.

Asylum for people with sensitive knowledge or expertise was routinely granted in the context of the Cold War and its aftermath. Section 7 of the \href{https://www.cia.gov/readingroom/docs/CIA-RDP89B00552R000700070018-7.pdf}{CIA Act of 1949} provided for admission and permanent residence of up to a hundred defectors and their immediate families per fiscal year if deemed “in the interest of national security or essential to the furtherance of the national intelligence mission.” The Soviet Scientists Immigration Act of 1992 gave up to 750 visas to former Soviet and Baltic States scientists with “expertise in nuclear, chemical, biological or other high technology fields or who are working on nuclear, chemical, biological or other high-technology defense projects.”

The challenge inspections mechanism we lay out in paragraph 3 of this article is modeled after that of Part IX of the \href{https://www.opcw.org/sites/default/files/documents/CWC/CWC_en.pdf}{CWC}:

\begin{displayquote}
Each State Party has the right to request an on-site challenge inspection of any facility or location in the territory or in any other place under the jurisdiction or control of any other State Party for the sole purpose of clarifying and resolving any questions concerning possible non-compliance…
\end{displayquote}

The CWC, along with other arms control treaties such as the INF and START I nuclear treaty between the U.S. and USSR, combines NTM with \href{https://www.osti.gov/servlets/purl/7166074}{challenge-like inspections} to verify compliance.

\paragraph{Notes on Article X}
\subparagraph{Intelligence Gathering}
We expect all parties would make ongoing efforts to independently determine whether any actor is conducting dangerous AI activities, out of interest in their own security. A range of state intelligence gathering activities would supplement and validate monitoring the CTB conducts directly (as described in Articles IV through VII). Towards that end, an Information Consolidation division is vital, and must be trustworthy to receive information from all parties. It will keep sensitive information confidential and secure, and must be sufficiently robust to assure state intelligence services that the risks imposed on their intelligence methods are minimal, and are justified in order to provide needed information to the CTB. Avoiding collecting sensitive information whenever possible, and keeping the collected information in the strictest confidence, minimizes risks of compromise.

Article X also addresses the surveillance of non-signatories, where the need for intelligence is strong.

Article X stops short of imposing an obligation to surveil. It would be unprecedented to mandate the creation of a self-sufficient intelligence gathering capability within the CTB at the required level of capability to give states assurance, and as such, it seems unnecessary in light of the fact that the creation of superintelligence would pose a grave security threat, which means all parties are already strongly incentivized to surveil and monitor any actor with that capability. Thus, the CTB relies primarily on parties to provide key intelligence.

\subparagraph{Whistleblower Protections}
The overall effectiveness of this agreement relies on parties’ justified confidence that other parties are not undertaking prohibited AI activities. Even with National Technical Means and other intelligence gathering, it may be difficult for states to detect clandestine efforts to develop superintelligence. There are many domains in which it may not be feasible for states to gather intelligence on their rivals, such as efforts conducted inside military facilities. Whistleblowers can serve as an additional source of information, and the possibility of whistleblowing provides further deterrence against non-compliance.

Whistleblowers may be effective because individuals involved in secret violations (e.g., clandestine training runs or AI research) may themselves be concerned about the danger from ASI. This article aims to make it safer and less costly for them to report violations, shifting the personal incentives away from silence and toward disclosure.

Whistleblowers could sound the alarm for violations including:
\begin{itemize}[nosep]
    \item Article IV: Training runs that are unmonitored, exceed thresholds, or use prohibited distributed training methods.
    \item Article V: The existence of undeclared chip clusters, the failure to consolidate all covered hardware, or the diversion of chips to secret, unmonitored facilities.
    \item Article VI: New manufactured AI chips diverted away from monitoring, or created without mandated security features.
    \item Article VIII: Prohibited AI research.
\end{itemize}

Modifications to the whistleblower clauses could change their efficacy and political viability in various ways. For example, states could offer to financially compensate legitimate whistleblowers to provide additional incentives, but this may be seen as paying citizens to defect on their own countries.

\subparagraph{Challenge Inspections}
Challenge inspections are a critical function provided by the agreement. Without the credible threat of detection, parties may fear that their rivals would attempt to cheat (despite the lose-lose nature of a race to superintelligence). Intelligence gathering is one method to combat apparent (illusory) incentives to defect.

\subparagraph{U.S. and PRC Independent Verification Efforts}
The existence of independent verification efforts provides redundancy and enhances the overall assurance that the Agreement is being implemented effectively. We believe the U.S. and the PRC will require independent verification founded on their own analysis, and the agreement and efforts of the CTB support these independent efforts as much as possible.

\subsection*{ARTICLE XI — Dispute Resolution}
\begin{enumerate}
    \item Any Party ("Concerned Party") may raise concerns regarding the implementation of this Agreement, including concerns about ambiguous situations or possible non-compliance by another Party ("Requested Party"). This includes misuse of Protective Actions (Article XII).
    \begin{enumerate}
        \item The Concerned Party notifies the Requested Party of their concern, while also sharing their concern with the Director-General and Executive Council. The Requested Party will acknowledge this notification within 36 hours, and provide clarification within 5 days.
    \end{enumerate}
    \item If the issue is not resolved, the Concerned Party may request that the Executive Council assist in adjudicating and clarifying the concern. This may include the Concerned Party requesting a challenge inspection in accordance with Article X.
    \begin{enumerate}
        \item The Executive Council provides appropriate information in its possession relevant to such a concern.
        \item The Executive Council may task the CTB to compile additional documentation, convene closed technical sessions, and recommend resolution measures.
    \end{enumerate}
    \item If the Executive Council determines there was a violation of the Agreement, it can take actions to prevent dangerous AI activities or reprimand the Requested Party. These actions may include:
    \begin{enumerate}
        \item Require additional monitoring or restrictions on AI activities
        \item Require relinquishment of AI hardware
        \item Call for sanctions
        \item Recommend Parties take Protective Actions under Article XII
    \end{enumerate}
\end{enumerate}

\paragraph{Precedent for Article XI}
Our Article XI Dispute Resolution procedures borrow from Articles IX, XII, and XIV of the \href{https://www.opcw.org/sites/default/files/documents/CWC/CWC_en.pdf}{Chemical Weapons Convention}. Article IX of the CWC requires signatories to respond to requests for clarification “as soon as possible, but in any case not later than 10 days after the request.” Given how quickly digital developments can propagate, we chose a 5-day response deadline, but even this figure may need to be adjusted downward.

Our paragraph 2 of this article is modeled after Article XIV of the CWC, which permits its Executive Council to “contribute to the settlement of a dispute by whatever means it deems appropriate, including offering its good offices, calling upon the States Parties to a dispute to start the settlement process of their choice and recommending a time-limit for any agreed procedure.” Parties are also encouraged to refer cases to the International Court of Justice as appropriate.

As in paragraph 3 of our Article XI, the CWC’s Article XII empowers the Executive Council to recommend remedies, including sanctions, “in cases where serious damage to the object and purpose of this Convention may result from activities prohibited under this Convention.” To give force to those recommendations, the CWC’s Council is to “bring the issue, including relevant information and conclusions, to the attention of the United Nations General Assembly and the United Nations Security Council.” Recommendations by our agreement’s Executive Council may be similarly escalated.

\paragraph{Notes on Article XI}
The purpose of Article XI is to include a consultation and clarification process to resolve issues that arise between signatories.

Given the pace of AI innovation, determining violations on a reasonable timeline can be challenging. The role of the Executive Council is to adjudicate any concerns raised by any party to the agreement. The CTB has the role of coordinating inspections by experts that have an understanding of cutting-edge AI technologies. The agreement uses an aggressive timeline (measured in hours and days) in the hopes that it is fast enough for parties to wait for rulings before taking Protective Actions (as described in Article XII, below), even despite the rapid pace of technological change in the field of AI. That said, of course no agreement can prevent a party from taking protective actions that they deem necessary to ensure their own security.

\subsection*{ARTICLE XII — Protective Actions}
\begin{enumerate}
    \item Recognizing that the development of ASI or other Dangerous AI Activities, as laid out in Articles IV through IX, would pose a threat to global security and to the life of all people, it may be necessary for Parties to this Agreement to take drastic actions to prevent such development. The Parties recognize that development of artificial superintelligence (ASI), anywhere on earth, would be a threat to all Parties. Under Article 51 of the United Nations Charter and as longstanding precedent, states have a right to self-defense. Due to the scale and speed of ASI-related threats, self-defense may require pre-emptive actions to prevent the development of ASI.
    \item To prevent the development or deployment of ASI, this Article authorizes tailored Protective Actions. Where there is credible evidence that a State or other actor (whether a Party or a non‑Party) is conducting or imminently intends to conduct activities aimed at developing or deploying ASI in violation of Article I, Article IV, Article V, Article VI, Article VII, or Article VIII, a State Party may undertake Protective Actions that are necessary and proportionate to prevent such activities. In recognition of the harms and escalatory nature of Protective Actions, Protective Actions should be used as a last resort. Outside of emergencies and time-sensitive situations, Protective Actions are preceded by other approaches such as, but not limited to:
    \begin{enumerate}
        \item Trade restrictions or economic sanctions
        \item Asset restrictions
        \item Visa bans
        \item Appeal to the UN Security Council for action
    \end{enumerate}
    \item Protective Actions may include measures such as cyber operations to sabotage AI development, interdiction or seizure of covered chip clusters, military actions to disable or destroy AI hardware, and physical disablement of specific facilities or assets directly enabling AI development.
    \item Parties minimize collateral harm, including to civilians and essential services, wherever practical, subject to mission requirements.
    \item Protective Actions are strictly limited to preventing ASI development or deployment and are not used as a pretext for territorial acquisition, regime change, resource extraction, or broader military objectives. Permanent occupation or annexation of territory is prohibited. Action will cease upon verification by the coalition that the threat no longer exists.
    \item Each Protective Action is accompanied, at initiation or as soon as security permits, by a public Protective Action Statement that:
    \begin{enumerate}
        \item Explains the protective purpose of the action;
        \item Identifies the specific AI‑enabling activities and assets targeted;
        \item States the conditions for cessation;
        \item Commits to cease operations once those conditions are met.
    \end{enumerate}
    \item Protective Actions terminate without delay upon any of the following:
    \begin{enumerate}
        \item Coalition certification that the relevant activities have ceased.
        \item Verified surrender or destruction of covered chip clusters or ASI‑enabling assets, potentially including the establishment of sufficient safeguards to prevent restricted research activities.
        \item A determination by the acting Party, communicated to the CTB, that the threat has abated.
    \end{enumerate}
    \item Parties do not regard measured Protective Actions taken by another Party under this Article as provocative acts, and do not undertake reprisals or sanctions on that basis. Parties agree that Protective Actions meeting the above requirements are not construed as an act of aggression or justification for the use of force.
    \item The Executive Council reviews each Protective Action for compliance with this Article. If the Executive Council finds that an action was not necessary, proportionate, or properly targeted, actions may be taken under Article XI, paragraph 3.
\end{enumerate}

\paragraph{Precedent for Article XII}
The idea that nation-states can take protective actions for their own security is a reality regardless of precedent, but one case of its codification into international law is \href{https://www.un.org/en/about-us/un-charter/chapter-7}{Chapter VII of the United Nations Charter}, which states that the Security Council may take military or non-military measures to maintain international peace and security, when necessary.

The concept of Protective Actions as they appear in the draft above is further grounded in historical precedents where states have acted, individually or collectively, to prevent the development of technologies deemed a threat to international security. These actions range from sanctions to cyber and military strikes.

The international effort to prevent Iran from developing nuclear weapons provides a clear, modern example. The UN Security Council has several times \href{https://www.cnn.com/2012/01/23/world/meast/iran-sanctions-facts/index.html}{imposed} sanctions on Iran due to its nuclear program, most of which were lifted after Iran agreed to limits on said program in the \href{https://main.un.org/securitycouncil/en/content/2231/background}{2015 Joint Comprehensive Plan of Action}.

The U.S. and Israel \href{https://www.nytimes.com/2012/06/01/world/middleeast/obama-ordered-wave-of-cyberattacks-against-iran.html}{reportedly collaborated} on Stuxnet, a highly sophisticated cyberweapon which destroyed many of Iran’s uranium enrichment centrifuges in 2010.

In \href{https://www.armscontrol.org/act/2025-07/news/israel-and-us-strike-irans-nuclear-program}{June 2025}, Israel launched airstrikes against many of Iran’s nuclear facilities, and this was followed by U.S. airstrikes nine days later which were partially aimed at disabling the Fordow Uranium Enrichment Plant.

Another historical precedent for Protective Actions is the international response to Iraq’s nuclear noncompliance in the 1990s. Following the 1991 Gulf War, the \href{https://www.un.org/depts/unscom/}{United Nations Special Commission} (UNSCOM) was created to oversee the destruction of Iraq’s weapons of mass destruction. Non-compliance with the UNSCOM inspection regime eventually led to \href{https://www.afhistory.af.mil/FAQs/Fact-Sheets/Article/458976/1998-operation-desert-fox/}{Operation Desert Fox} in 1998, a bombing campaign aimed at degrading Iraq’s ability to produce WMDs.

\paragraph{Notes on Article XII}
An agreement to prevent the creation of artificial superintelligence might not need to be explicit about the need for Protective Actions against states undertaking ASI development, and instead leave these dynamics implicit, as similar agreements often do. Our draft is explicit because this deterrence regime is core to the effectiveness of the agreement, and clarity around the incentives increases the effectiveness. This explicitness also allows us to include measures that may help prevent Protective Actions being misused, including more thorough description of when these Actions are acceptable.

Once world leaders understand the threat from ASI, they will likely be willing to take action to stop rogue AI development, including limited military interventions. Military actions, such as narrowly targeted airstrikes, should always be treated as a last resort option to prevent the development of ASI, after all other diplomacy has failed. But it is important that they are available as a last resort, in order for the deterrence and compliance regime to hold even towards actors who wrongly perceive recklessly created artificial superintelligence as a technology that would be beneficial rather than destructive.

We stress that any use of force should be targeted at preventing ASI, and should stop once it is clear that the threat has been removed. Article XII aims to make it clear that signatories would not prevent reasonable Protective Actions taken by other parties, but these actions must also be reviewed to ensure that this article is not being abused.

\subsection*{ARTICLE XIII — Coalition Technical Body Reviews}
\begin{enumerate}
    \item For AI models created via declared training or post‑training within the limits of Article IV, the CTB may require evaluations and other tests. These tests will inform whether the thresholds set in Article IV, Article V, Article VII, and Article VIII need to be revised. The methods used for reviews will be determined by the CTB and may be updated.
    \item Evaluations are conducted at CTB facilities or monitored CCCs, by CTB officials. Officials from Parties to the Agreement may be informed which tests are conducted, and the CTB may provide a summary of the test results. Parties will not gain access to AI models they did not train, except when granted access by the model owner, and the CTB will take steps to ensure the security of sensitive information.
    \item The CTB may share detailed information with Parties or the public, if the Director-General deems that this may be necessary to reduce the chance of human extinction from advanced AI.
\end{enumerate}

\paragraph{Precedent for Article XIII}
Precedents for tests with oversight are shared with precedents around chip use verification discussed under Article VII, with the missile telemetry sharing protocol of START I being particularly relevant. The added component here in our Article XIII is using collected data to inform recommendations for potential threshold adjustments (which could take place under the precedented mechanisms we discuss with Article XIV).

Regarding the inherent tension between disclosures to the public (paragraph 3) and the information consolidation provisions of our Article X, we note that the \href{https://www.iaea.org/about/statute}{Statute of the IAEA}’s Article VII confidentiality provision\footnote{VII.F states that “[...] subject to their responsibilities to the Agency, [the Director General and the staff] shall not disclose any industrial secret or other confidential information coming to their knowledge by reason of their official duties for the Agency”} has not prevented it from publishing \href{https://www.iaea.org/publications/reports}{regular and detailed reports} on major developments in its associated field and their implications for global security.

\paragraph{Notes on Article XIII}
The purpose of Article XIII is to ensure the CTB stays up to date with the state of the field of AI, in case it is advancing. For example, reviewing declared training would allow it to understand the level of AI capabilities that can be reached with different levels of training FLOP. Even with algorithmic research prohibited, there may be progress that cannot be effectively stopped, and the CTB must keep track of it.

Additionally, the CTB has reason to monitor progress in capabilities elicitation. For example, new prompting methods could be discovered that cause an old AI to perform much better on some critical evaluation metric.

We envision CTB reviews that also involve capability evaluations to make sure AIs aren’t getting dangerously capable in specific domains. They could also look at the training data to ensure AIs aren’t being trained for specifically dangerous tasks (like automating AI research), or to test for unexpected AI behavior.

When reviews reveal shifts in the AI development landscape, those shifts could necessitate changes to thresholds relevant to Article IV and Article V, and changes to the definitions of restricted research in Article VIII, with those changes implemented according to the mechanisms in Article III.

\subsection*{ARTICLE XIV — Revision Process}
\begin{enumerate}
    \item The Executive Council may revise this Agreement as necessary to ensure its purposes are achieved. "Amendments" are considered revisions to the main body and Articles of the Agreement. Under Article III, the CTB may change specific definitions and implementation methods, such as those relevant to Article IV, Article V, Article VI, Article VII, Article VIII, Article IX, and Article X, subject to the Executive Council's veto power. Fundamental revisions to the purposes of these Articles or to the governance structure require an Amendment by the Executive Council.
    \item The Executive Council may propose amendments to all Parties to the Agreement. The Executive Council shall circulate proposed amendments to all States Parties with an explanation of the rationale and expected effects.
    \item Parties to the Agreement may submit recommendations for amendments to the Executive Council through the Director-General. The Executive Council will consider such recommendations but is not obligated to adopt them.
    \item Amendments proposed by the Executive Council become effective upon consensus of the Executive Council.
    \item Three years after the entry into force of this Agreement, the Executive Council shall convene a review conference to assess the operation of this Agreement with a view to assuring that the purposes of the Preamble and the provisions of the Agreement are being realized. All Parties to the Agreement shall be invited to participate. At intervals of three years thereafter, the Executive Council will convene further review conferences with the same objective.
\end{enumerate}

\paragraph{Precedent for Article XIV}
Our agreement is resilient to short-term pressures to relax thresholds or weaken provisions, as it requires consensus by members of the Executive Council to make such changes.

Hard-to-amend (and thus hard-to-weaken) treaties rely on other mechanisms for strengthening as needed. The NPT has never been amended, but has been adapted through the five-yearly Review Conference stipulated in Article VIII, where consensus agreements are made “with a view to assuring that the purposes of the Preamble and the provisions of the Treaty are being realised.”

Similarly, Article XII of the 1975 \href{https://treaties.unoda.org/t/bwc}{Biological Weapons Convention} relies on its five-yearly Review Conferences to strengthen the treaty through non-binding Confidence-Building Measures, as formal amendments are rare. Our agreement stipulates a three-year conference, as AI has been a field prone to rapid shifts; this period may need to be further shortened.

Article XV of the \href{https://www.opcw.org/sites/default/files/documents/CWC/CWC_en.pdf}{Chemical Weapons Convention} makes a distinction between amendments and administrative or technical changes, with less stringent approval provisions for the latter. Similar language could be added to our draft agreement to provide a level of flexibility in managing future developments in the field of AI.

Article XV of the \href{https://www.unoosa.org/oosa/en/ourwork/spacelaw/treaties/outerspacetreaty.html}{Outer Space Treaty} contains an amendment clause, but the treaty has never been formally amended; instead, new treaties have been negotiated to address emerging space issues. This could be another option for shoring up weaknesses that may become apparent in an AI agreement.

\paragraph{Notes on Article XIV}
Article XIV sets out the process to make major revisions to the agreement. These revisions require substantial support from the parties and there is a high bar to make such revisions. By contrast, changes to the details of various categories and restrictions can be made much more easily and rapidly (subject to slower review), as described in Article III, and as is necessitated by the fast pace of change in the field of AI. A careful review process seems warranted given the gravity of the situation, and given the risk that overzealous actors could, if left unchecked, impose misguided restrictions that inconvenience the public for little-to-no benefit.

\subsection*{ARTICLE XV — Withdrawal and Duration}
\begin{enumerate}
    \item The Agreement shall be of unlimited duration.
    \item Each Party will, in exercising its national sovereignty, have the right to withdraw from the Agreement if it decides that extraordinary events, related to the subject matter of this Agreement, have jeopardized the supreme interests of its country. It shall give notice of such withdrawal to the CTB 12 months in advance.
    \item During this 12-month period, the withdrawing state shall cooperate with CTB and Executive Council member efforts to certify that after withdrawal, the withdrawing state will be unable to develop, train, post-train, or deploy dangerous AI systems, including ASI or systems above the Agreement thresholds. Withdrawing states acknowledge that such cooperation aids the CTB and Parties in avoiding the use of Article XII.
    \begin{enumerate}
        \item In particular, the withdrawing state, under CTB and Executive Council member oversight, will remove all covered chip clusters and ASI-enabling assets (e.g., advanced computer chip manufacturing equipment) from its territory to Executive Council-approved control or render them permanently inoperable (as described in Article V).
    \end{enumerate}
    \item Nothing in this Article limits the applicability of Article XII. A State that has withdrawn (and is therefore a non-Party) remains subject to Protective Actions if credible evidence indicates activities aimed at ASI development or deployment.
\end{enumerate}

\paragraph{Precedent for Article XV}
It is common for treaties to lack expiration dates. The first paragraph of \href{https://www.opcw.org/chemical-weapons-convention/articles/article-xvi-duration-and-withdrawal}{Article XVI of the CWC} states “This Convention shall be of unlimited duration.”

Agreements of unlimited duration do not necessarily last forever.\footnote{Sometimes they are superseded by other treaties. This was the case for the 1947 \href{https://www.wto.org/english/docs_e/legal_e/gatt47_e.htm}{General Agreement on Tariffs and Trade} (GATT); it was superseded by the 1994 \href{https://www.wto.org/english/docs_e/legal_e/marag_e.htm}{Marrakesh agreement}, which incorporated the rules from GATT but established the World Trade Organization (WTO) to replace GATT’s institutional structure. Treaties of unlimited duration also sometimes end when parties withdraw in a manner that makes the treaty ineffective. For example, the U.S. and USSR initially agreed to the 1987 \href{https://2009-2017.state.gov/t/avc/trty/102360.htm}{Intermediate-Range Nuclear Forces (INF) Treaty} for an unlimited duration, but the U.S. withdrew in 2019 citing Russian non-compliance, and Russia later announced it would no longer abide by the treaty in 2025.} But they do typically provide a mechanism for withdrawal, usually with a required period of notice and other stipulations that might let it leave in a manner less concerning to the remaining parties. \href{https://www.opcw.org/chemical-weapons-convention/articles/article-xvi-duration-and-withdrawal}{Article XVI of the CWC} allows for a party to withdraw “if it decides that extraordinary events, related to the subject-matter of this Convention, have jeopardized the supreme interests of its Country.” The withdrawing country must give 90 days notice. \href{https://www.unoosa.org/oosa/en/ourwork/spacelaw/treaties/outerspacetreaty.html}{Article XVI of the Outer Space Treaty} requires one year notice for withdrawal.

Our agreement expects 12 months' notice from departees, allowing ample time for assisting with the assurance-providing measures in Paragraph 3. Our intent with these measures (which go beyond what we readily find in the historical record of withdrawal provisions) is to reduce the potential need for protective actions against the withdrawing Party, as no Party or non-Party can be allowed to create ASI or weaken the world’s ability to prevent its creation.

Historical precedent for a withdrawn party remaining subject to protective actions is found in the case of \href{https://main.un.org/securitycouncil/en/s/res/1718-\%282006\%29}{United Nations Security Council Resolution 1718}, which imposed sanctions against North Korea after its 2006 nuclear test, despite North Korea’s previous withdrawal from the NPT.

\paragraph{Notes on Article XV}
Given the dangers of ASI research and development, as well as the risk that if one country decides to withdraw from the agreement and race to superintelligence then others might follow, the agreement needs barriers to withdrawal.

In practice, this is challenging. North Korea, for example, withdrew from the NPT to continue its nuclear proliferation activities, even at the cost of UN Security Council resolutions and associated sanctions. The consequences did not prove sufficient to cause North Korea to stop its proliferation activities.

If nations wish to withdraw from the agreement, our wording makes it clear that, in the eyes of all parties, they forgo the right to AI infrastructure, and that they would be subject to Article XII Protective Actions if they engage in dangerous AI activities. Any further negotiation around the ASI issue — e.g., to avoid Protective Actions — would have to be negotiated separately by interested parties.

Parties concerned about withdrawals could include mechanisms to make withdrawal more difficult. For example, both U.S. and Chinese officials could agree to install mutual killswitches inside covered chip clusters, allowing either party to permanently shut off the other’s clusters. Alternatively, parties could adopt a multilateral licensing regime in which all new AI chips must be fabricated with hardware locks that require approval from multiple parties to continue operation, so that if a country withdrew from the agreement, others could stop approving their licenses and incapacitate their chips. Another option involves moving key AI infrastructure into third-party countries where the infrastructure could be confiscated or destroyed if a party withdrew from the agreement. Our draft sticks to minimal deterrence methods, but many other methods are available (or could be made available with a little technological investment).

\section{What can we do today?}
\label{app:whatcanwedo}
Recognizing that measures to address the risks from AI cannot be developed overnight, we are providing a list of measures that could be implemented, beginning today, that would help lay the foundation for mitigating risks from AI. Our intention is to provide a list of actions that we believe will be necessary to build a robust framework and basis for future AI agreements and risk reduction measures. It is not a comprehensive list, or a detailed explanation of each measure, but rather a starting point for discussion.

These measures fall into three categories:

\begin{enumerate}
    \item \textbf{Preparing for an International Agreement:}
    \begin{itemize}[nosep]
        \item Develop situational awareness of where AI chips are located globally, identify public and nonpublic data centers, understand chip and/or hardware smuggling pipelines, and production flows around the world. Begin tracking AI chips.
        \begin{itemize}[nosep]
            \item Establish robust AI chip, hardware security, and supply chain standards internationally.
            \begin{itemize}[nosep]
                \item Conduct R\&D for on-chip hardware-enabled governance mechanisms (HEMs) and for supporting equipment that could verify the location and the types of workloads conducted on advanced AI chips (e.g., training, inference). Develop tamper-resistant HEM technologies and explore advanced HEM capabilities that allow for remote governance, licensing, and limits on chip use.
                \item Establish an interagency group within the U.S. government to rapidly develop the framework for implementing location verification and advanced HEMs.
                \item Establish a public/private partnership to develop advanced HEM technologies
                \item Set up “governing through the cloud” \citep{heim_governing_2024} style approaches and enhanced capacity for overall chip use governance.
                \item Ensure export control measures are flexible enough to adjust to the state of the art.
            \end{itemize}
        \end{itemize}
        \item Ensure advanced AI chip development, hardware, and supply chains remain in a small group of countries.
        \item Enhance international coordination on controls to ensure AI chips, R\&D, hardware, and tacit knowledge are not easily obtained by groups of concern.
        \item Back proactive measures to implement future AI governance and an overall agreement and avoid actions that make an international agreement more difficult, like building covert data centers.
        \item Establish international multi-disciplinary collaboration among policymakers, technologists, and institutions to establish the foundation of future international discussions on AI governance.
        \item Establish AI hotlines between key actors in the U.S. and PRC government.
        \item Work with the PRC to establish open-model-weight and evaluation standards related to CBRN risks, e.g. biological weapon development.
        \item Work with countries with sizable AI capabilities to share large data center locations.
    \end{itemize}

    \item \textbf{Building the Capacity of Future Governance, Safety, and Security}
    \begin{itemize}[nosep]
        \item Invest in education and recruiting talent to focus on AI monitoring, threat evaluation, and safety research.
        \item Fund research focused on AI evaluations (both before and after model deployment), safety research measures to mitigate risk, and trigger points for implementation of such measures.
        \item Invest in verification expertise and know-how, for instance by funding pilot verification efforts using open-source intelligence or satellite data.
        \item Identify gaps between current expertise and expertise needed in the future to ensure a strong security/safety ecosystem for research, implementation, and cross-pollination.
        \item Identify experts in adjacent fields that could be brought to bear on AI risks.
        \item Implement public and/or private NGO initiatives to create expertise where current gaps exist.
        \item Identify current areas of consensus, and build up to more robust measures to address the risks of AI.
        \item Enhance security for current and future AI projects, in order to protect against proliferation to and misuse by third parties including terrorists or rogue states.
        \item Regulate agentic AI interactions as the capabilities come online.
        \item Increase AI fluency in the general public to ensure awareness of implications of AI development.
    \end{itemize}

    \item \textbf{Domestic Measures that Could be Implemented Today:}
    \begin{itemize}[nosep]
        \item Establish robust non-public information sharing between AI labs and key government officials to ensure the government is privy to all important state-of-the-art AI developments that could have an impact on the economy, national security, or threat environment.
        \item Establish a cross-functional team within the U.S. government for AI model assessment, and develop a standardized assessment framework for review of existing and planned domestic and international AI models. This group should analyze the development of current and future AI, identify key AI developmental milestones, and develop risk mitigation strategies before specific milestones are reached.
        \item Centralize AI strategy development in a White House-led organization either within OSTP or as a U.S. government-wide steering committee to identify gaps and harmonize strategy/implementation.
        \item Task an executive branch agency with assessing the implications of advanced AI on key national security areas including CBRN and cyber capabilities, model autonomy, and automated AI R\&D to speed up AI developments, and open-source models.
        \item Develop scenario-based incident response capabilities \citep{somani2025strengthening}.
    \end{itemize}
\end{enumerate}

\section{Stages}
\label{app:stages}
International discussions on AI governance and the establishment of guardrails remain at a nascent stage. Predicting the path such negotiations might take is inherently difficult, but this section illustrates \textit{one possible pathway} to governance—one that begins with bilateral measures between the U.S. and PRC and gradually expands to involve relevant states. This should be considered a hypothetical scenario rather than a recommendation.

The purpose of this section is to demonstrate how AI governance discussions could evolve in practice. It takes into account the stages of how transparency and confidence-building measures and agreements have been agreed to in other historical contexts.

\paragraph{Stage One – Laying the Foundation: Initial Transparency and Confidence Building}
\textit{Build initial capacity, foster collaboration, and establish foundational transparency mechanisms.}
\begin{itemize}[nosep]
    \item The \textbf{U.S. and PRC issue a joint declaration} at the head of state level on responsible AI development and deployment. They commit to:
    \begin{itemize}[nosep]
        \item Avoid AI use for certain prohibited  purposes
        \item A “No First Use” policy for AI-driven cyberattacks targeting critical infrastructure, including power grids, financial systems, healthcare networks, and election systems. (This builds upon the 2024 U.S.–PRC agreement restricting AI control over nuclear weapons systems and reduces the risk of sudden or destabilizing strategic attacks)
    \end{itemize}
    \item \textbf{AI Transparency, Monitoring, and Information Sharing}
    \begin{itemize}[nosep]
        \item U.S. and PRC, and possibly other like-minded countries begin monitoring AI developments and sharing relevant information
        \item Conduct limited public disclosures to highlight emerging risks
        \item Domestic and International Tracking
        \begin{itemize}[nosep]
            \item Track AI hardware, chips, and AI research for future verification mechanisms
            \item Make limited declarations of such data
        \end{itemize}
    \end{itemize}
    \item \textbf{R\&D for Verification}
    \begin{itemize}[nosep]
        \item Begin early research and development on verification methods that could serve future governance mechanisms
    \end{itemize}
    \item \textbf{Coalition Building}
    \begin{itemize}[nosep]
        \item Launch initial conversations among interested governments and stakeholders 
        \item Establish governance concepts for future international AI governance
        \item Allocate resources to support international AI-safety research and coordination
    \end{itemize}
\end{itemize}

\paragraph{Stage Two – Enhanced Transparency and Communication}
\textit{These measures build confidence and transparency and reduce the risk of misunderstanding rapid AI developments as signs of imminent aggression.}
\begin{itemize}[nosep]
    \item Establish \textbf{secure communication channels} between high-level AI and cybersecurity officials in the U.S. and PRC
    \item Exchange of information between the U.S. and PRC: Conduct \textbf{annual confidential disclosures on high-consequence AI systems} (excluding model weights or algorithms), including:
    \begin{itemize}[nosep]
        \item Purpose and capabilities of each system
        \item Hardware used for operation
        \item Data center locations and capacities
    \end{itemize}
    \item U.S. and PRC \textbf{pre-notify} of major AI model \textbf{training runs} and deployments exceeding agreed thresholds
    \item \textbf{Develop basis for future international AI governance}
    \begin{itemize}[nosep]
        \item Make multilateral declarations defining safe AI development
        \item Develop voluntary commitments that enhance safe AI development practices and limit use cases that are escalatory or unsafe
        \item Ongoing investment in international AI safety research and viable prototype verification mechanisms
        \item Establish pooled resource mechanisms, allowing nations to benefit from AI advancements while obviating the need for fully independent programs with associated costs
        \item Begin voluntary information disclosures among additional nations:
        \begin{itemize}[nosep]
            \item Number of data centers and large-scale models that exceed agreed thresholds
            \item Bilateral disclosure mechanisms serve as templates for multilateral transparency frameworks, allowing other countries to join using standardized reporting formats
        \end{itemize}
    \end{itemize}
\end{itemize}

\paragraph{Stage Three – Establishing Commitments, Limits, and Verification Foundations}
\textit{Move from voluntary transparency to structured commitments and verified limitations. Verification systems and commitments piloted bilaterally evolve into the technical and procedural framework for broader multilateral agreements.}
\begin{itemize}[nosep]
    \item U.S. and PRC lead discussions and negotiations on AI governance measures including: a ban on AI training above the Strict Threshold; chip consolidation; chip production monitoring; chip use verification; research restrictions and verification detailed in Articles IV, V, VI, VII, VIII, IX and other articles included in the Agreement in Appendix~\ref{app:agreement}
    \item The U.S. and PRC establish and test chip use verification protocols to confirm future compliance within the limit mentioned in the first bullet
    \item The U.S., PRC, and additional countries make safety and development commitments and invest in shared research infrastructure
\end{itemize}

\paragraph{Stage Four – Institutionalization and Deployment}
\textit{Translate practical cooperation into a formalized international agreement.}
\begin{itemize}[nosep]
    \item The measures of the international agreement are implemented. (See Appendix~\ref{app:agreement})
\end{itemize}

\paragraph{Stage Five – Solving for ASI}
\textit{Make use of the time and resources available to prepare for ASI}
\begin{itemize}[nosep]
    \item The world directs capacity toward solving the problems required to permit the safe development of superintelligence. These likely include AI alignment research, managing social, economic, and geopolitical destabilization, preventing human misuse of advanced AI, and avoiding concentration of power, but this list could look different when more is known about ASI development
    \item Establish joint resilience mechanisms for critical infrastructure
    \item Build global safety coordination systems capable of rapid response to emerging risks
    \item Maintain continuous cooperation between the Executive Council and national institutions
    \item Adapt the agreement in response to a changing geopolitical environment or potentially a changing technical environment around AI development
\end{itemize}

\paragraph{Stage Six – Safe implementation of superintelligence}
\begin{itemize}[nosep]
    \item The world implements the solutions discovered above and maintains safety and coordination mechanisms
\end{itemize}

This illustrates a possible path to a framework for AI governance, progressing from bilateral commitments and transparency to global cooperation. Early U.S.–PRC engagement acts as the catalyst for broader international participation, with each stage reinforcing the subsequent stage.

\section{Locating and consolidating AI chips}
\label{app:locatingchips}
The consolidation of AI chips into monitored facilities is a central part of the agreement. Success requires both domestic authorities to locate and secure all relevant compute infrastructure within their borders, and international verification mechanisms to ensure these processes are conducted comprehensively and in good faith.

Article V of the agreement (Appendix~\ref{app:agreement}) designates all clusters which must be monitored as covered chip clusters (CCCs). In our plan, a CCC is defined as a networked cluster or any set of AI chips with aggregate effective computing capacity greater than 16 H100-equivalents. The credibility of the governance framework rests on the coalition's ability to account for and monitor all covered chip clusters (CCCs).

\subsection{Key goals}
The chip consolidation process must solve two key problems:

\paragraph{Domestic Discovery and Consolidation.} Each country must use domestic capabilities (or optionally enlist external assistance) to locate all CCCs within its borders. This requires not only identifying known commercial facilities but also detecting any unregistered or concealed CCCs that might be operated by smaller organizations and individuals.

\paragraph{International Verification.} Other coalition members must have confidence that domestic authorities have not accidentally overlooked any CCCs. More critically, they must trust that states have not deliberately concealed CCCs or diverted chips into secret facilities for prohibited training runs. This concern is particularly acute between major powers: the U.S. must be able to verify that the PRC has declared all relevant compute infrastructure, and vice versa. Without this mutual confidence, the governance framework risks collapse due to suspicion of covert development.

To address this, states should use their existing intelligence capabilities and voluntary information-sharing mechanisms before the agreement enters into force. Pre-agreement intelligence gathering can establish baseline confidence about the location and scale of compute clusters in other jurisdictions.

\subsection{Consolidated chip ownership and operation}
The consolidation plan aims to maintain flexibility in ownership while ensuring adequate monitoring. Individuals and organizations can retain ownership of their AI chips and may continue to use them for permitted activities (e.g., inference). But these chips must be housed in facilities that meet monitoring and security requirements.

Existing large data centers (i.e., data centers housing >10,000 H100-equivalents) can be converted into monitored facilities, provided they can be retrofitted with the necessary security and monitoring infrastructure. This will require installation of tamper-evident/proof mechanisms and physical security measures to prevent unauthorized access or chip diversion. For facilities that cannot be adequately secured or monitored, the chips must be relocated to different facilities.

Smaller clusters (i.e., <1,000 H100-equivalents) must be consolidated into a manageable number of facilities that can be properly secured and continuously monitored.

\subsection{Methods for chip tracking and consolidation}
There are various levers available for both the chip consolidation process and the international verification process. There will be some crossover between these categories. Table~\ref{tab:domestic-chips} and Table~\ref{tab:international-chips} summarize which levers are useful for the domestic consolidation and international verification efforts, respectively. 

\subparagraph{Mandatory reporting of CCCs}
Mandatory reporting is a foundation for chip consolidation efforts. All individuals and organizations with clusters above the CCC threshold must declare their compute resources to domestic authorities. This mechanism should be easily accessible to domestic authorities. The international verification effort will likely also require access to this reported compute.

\subparagraph{Supply Chain Tracking}
Supply chain tracking can help verify that mandatory reporting has captured all relevant compute resources. It will mainly focus on obtaining sales records from AI chip manufacturers and distributors covering the previous few years. This includes major GPU manufacturers (NVIDIA, AMD, Intel, Huawei), specialized AI chip startups (Cerebras, Groq, and others), and Server OEMs and rack vendors (Dell Technologies, HPE, Supermicro, Lenovo). These records should detail which entities purchased chips, allowing for the supply chain to be tracked. Data for academic purchases of AI hardware will also be readily available to the domestic authorities.

This information serves as a primary source for both domestic authorities and international verification efforts. Domestic authorities gain access to complete records, enabling them to cross-reference declared clusters against actual chip distributions. For international verification, this data provides some assurance against state-run secret projects; even classified government programs will likely have sourced their chips through traceable channels at some point in the supply chain. International inspectors could conduct spot checks on limited samples of sales data rather than reviewing complete records.

\subparagraph{Inspections}
Physical inspections involve sending inspectors into facilities to check for AI chips. Large data centers contain extremely obvious and specialized compute infrastructure (server racks, cooling equipment, and other distinctive features). From inside a data center, it will be immediately clear that it is a data center.

Many of the other verification methods can identify sites that may potentially house AI chips. Inspections can then be used to confirm or disprove these suspicions. This will be a necessary part of both domestic and international verification efforts. Once a site is suspected of having a large number of chips, personnel need to physically verify the presence (or absence) and scale of compute infrastructure.

\subparagraph{Storage and Compensation Programs}
Governments may offer compensation for individuals or organizations to proactively consolidate their chips. There are multiple participation models: governments can purchase chips outright for placement in CCCs, or owners can retain control while housing their compute in monitored facilities. Inducements such as subsidized electricity can help compliance. Broken AI chips may be purchased by governments at a reduced cost, in order to certify their destruction and to ensure these chips are not repaired and then used in an unmonitored facility.

These programs could account for a large fraction of the total domestic consolidation effort. Inventory data can also be shared with international verification efforts.

\subparagraph{Satellite and Aerial Surveillance}
Large data centers are visible with satellite imagery and aerial surveillance. Facilities housing more than 10,000 H100-equivalents typically occupy substantial footprints with characteristic features including electrical substations. Due to the large amount of power and waste heat that AI chips produce, these sites also generally have distinctive thermal signatures.

Remote sensing primarily serves international verification efforts, providing independent validation of declared facilities without requiring ground access. While less precise than on-site inspection, satellite and aerial surveillance offers continuous, non-intrusive monitoring that can detect major discrepancies between declared and actual infrastructure. Domestic authorities can also leverage these capabilities to identify potentially undeclared facilities, though their primary verification tools typically have higher resolution.

\subparagraph{Power Consumption Monitoring}
One H100 GPU has a power consumption of 700 W, similar to that of the average American home. A data center with hundreds of thousands of chips has similar power use to a small city. Facilities with this sort of power draw require specialized electricity infrastructure, such as large transmission lines. This electricity infrastructure can easily be seen via satellite; even publicly available satellite imagery reveals transmission lines for facilities like aluminum smelters and desalination plants which have comparable power requirements.

It is possible to bury transmission lines, but this construction effort would take a long time and would itself be visible to satellites. It seems unlikely that there are currently large data centers powered by buried transmission lines.

Once a facility is determined to be using anomalously high power, it should be straightforward for domestic authorities or the international verification effort to perform an in-person inspection. This might involve sending inspectors to all aluminum smelters to ensure they have not secretly been refurbished into clandestine data centers.

Electricity data is generally less sensitive than other intelligence sources, making it more readily shareable for the international verification effort.

\subparagraph{Human Intelligence and Interviews}
The establishment and operation of data centers involves extensive human networks: AI researchers, data center operators, HVAC technicians, hardware vendors, and many more. Interviews with these individuals can reveal undeclared facilities or discrepancies in reported configurations.

For domestic efforts, interviews provide validation of other intelligence sources and can help with tracking of AI supply chains and data center construction. International verification efforts could conduct more limited interview programs, primarily as spot checks to confirm domestic reporting.

\subparagraph{Firmware-Based Chip Tracking}
Modern AI hardware includes security features such as secure enclaves that could enable location tracking and usage verification through firmware updates. While not robust against determined nation-state adversaries, firmware-based tracking provides an additional layer of security particularly effective against non-state actors attempting to evade monitoring.

There are some practical challenges which mean firmware-based tracking cannot be wholly relied on. Current chips do not have such tracking firmware, and so would need to be updated to allow tracking. However, chip updates would be voluntary, and individuals could simply choose to not update their chips. Additionally, the security of current AI chips is likely not robust to nation-state efforts.

These firmware updates could be bundled with essential security patches to incentivize adoption, though this approach would primarily help with individuals rather than state-sponsored projects. Both domestic and international verification efforts could leverage this data, though its limitations restrict its role to supplementary verification.

\subparagraph{Sting Operations}
Sting operations can identify actors attempting to establish unmonitored compute clusters or trade in prohibited unmonitored AI hardware. These operations might involve domestic authorities pretending to sell or purchase AI chips, or otherwise infiltrating smuggling networks.

Such operations are mainly useful for the domestic consolidation process, as cross-border sting operations raise significant sovereignty concerns. Information from these domestic operations may be shared with the international verification effort. While likely not applicable to state-run violations, these sting operations can prevent the accumulation of unmonitored compute by non-state actors that might otherwise enable prohibited training runs.

\subparagraph{Black Market Monitoring}
The existing number of AI chips distributed via smuggling networks will require dedicated tracking efforts. There have likely already been 100,000 to 1 million H100s smuggled into the PRC. This black market monitoring should encompass the chip repair sector, smuggling networks, and informal redistribution channels that could aggregate chips into unmonitored clusters.

This intelligence is valuable for both domestic and international verification. Domestic authorities need visibility into black market activity to prevent cluster formation outside the monitoring framework. International verification particularly values this intelligence to ensure that state projects are not sourcing chips through unofficial channels, circumventing supply chain tracking.

\subparagraph{Whistleblower Programs}
Voluntary reporting mechanisms for whistleblowers can provide useful intelligence on hidden chip stockpiles and black market activity. States will likely only enact an international agreement similar to what this report proposes in scenarios where there is widespread concern about AI risk, and so in these scenarios there will likely be whistleblowers.

National leaders can improve the effectiveness of whistleblower programs by promoting the reporting of concealed compute as a patriotic or humanitarian duty. While domestic programs can use whistleblower programs, international verification benefits most from these efforts, as they provide intelligence otherwise inaccessible to foreign inspectors. Even carefully compartmentalized state projects will likely involve many personnel, creating opportunities for disclosure even if individuals are selected for discretion.
\begin{table}[h!]
\centering
\caption{Domestic round-up. \\ \small\textit{Key: $\bullet$ = very useful for this purpose; $\circ$ = somewhat useful, but can't be relied on.}}
\label{tab:domestic-chips}
\resizebox{\textwidth}{!}{%
\begin{tabular}{|l|c|c|c|c|c|c|}
\hline
\textbf{\#H100-equivalents} & \textbf{100,000} & \textbf{10,000} & \textbf{1,000} & \textbf{100} & \textbf{16} & \textbf{1} \\ \hline
Seconds to Monitored Threshold & $1.0 \times 10^2$ & $1.0 \times 10^3$ & $1.0 \times 10^4$ & $1.0 \times 10^5$ & $6.3 \times 10^5$ & $1.0 \times 10^7$ \\ \hline
Seconds to Strict Threshold & $1.0 \times 10^4$ & $1.0 \times 10^5$ & $1.0 \times 10^6$ & $1.0 \times 10^7$ & $6.3 \times 10^7$ & $1.0 \times 10^9$ \\ \hline
Days to Monitored Threshold & 0.001 & 0.012 & 0.1 & 1.2 & 7.3 & 116.9 \\ \hline
Days to Strict Threshold & 0.1 & 1.2 & 11.7 & 116.9 & 730.7 & 11691.0 \\ \hline
Days to register clusters & Day 1 & Day 1 & Day 10 & Day 100 & Year 2 & Never \\ \hline
\hline
Mandatory reporting of CCCs & $\bullet$ & $\bullet$ & $\bullet$ & $\bullet$ & $\bullet$ &  \\ \hline
Supply chain tracking & $\bullet$ & $\bullet$ & $\bullet$ & $\circ$ & $\circ$ &  \\ \hline
Inspections & $\bullet$ & $\bullet$ & $\bullet$ & $\bullet$ & $\circ$ &  \\ \hline
Storage and compensation programs &  &  & $\circ$ & $\bullet$ & $\bullet$ &  \\ \hline
Satellite and aerial surveillance & $\bullet$ & $\bullet$ & $\circ$ &  &  &  \\ \hline
Electrical Power Monitoring & $\circ$ & $\circ$ &  &  &  &  \\ \hline
Human intelligence and interviews & $\bullet$ & $\bullet$ & $\bullet$ & $\bullet$ & $\circ$ &  \\ \hline
Firmware-based chip tracking &  &  &  & $\circ$ & $\circ$ &  \\ \hline
Sting operations &  &  &  & $\bullet$ & $\bullet$ &  \\ \hline
Black market monitoring &  &  & $\circ$ & $\bullet$ & $\bullet$ &  \\ \hline
Whistleblower programs & $\circ$ & $\circ$ & $\circ$ & $\circ$ & $\circ$ &  \\ \hline
\end{tabular}%
}
\end{table}

\begin{table}[h!]
\centering
\caption{International verification. \\ \small\textit{Key: $\bullet$ = very useful for this purpose; $\circ$ = somewhat useful, but can't be relied on.}}
\label{tab:international-chips}
\resizebox{\textwidth}{!}{%
\begin{tabular}{|l|c|c|c|c|c|c|}
\hline
\textbf{\#H100-equivalents} & \textbf{100,000} & \textbf{10,000} & \textbf{1,000} & \textbf{100} & \textbf{16} & \textbf{1} \\ \hline
Seconds to Monitored Threshold & $1.0 \times 10^2$ & $1.0 \times 10^3$ & $1.0 \times 10^4$ & $1.0 \times 10^5$ & $6.3 \times 10^5$ & $1.0 \times 10^7$ \\ \hline
Seconds to Strict Threshold & $1.0 \times 10^4$ & $1.0 \times 10^5$ & $1.0 \times 10^6$ & $1.0 \times 10^7$ & $6.3 \times 10^7$ & $1.0 \times 10^9$ \\ \hline
Days to Monitored Threshold & 0.001 & 0.012 & 0.1 & 1.2 & 7.3 & 116.9 \\ \hline
Days to Strict Threshold & 0.1 & 1.2 & 11.7 & 116.9 & 730.7 & 11691.0 \\ \hline
Days to register clusters & Day 1 & Day 1 & Day 10 & Day 100 & Year 2 & Never \\ \hline
\hline
Mandatory reporting of CCCs & $\bullet$ & $\bullet$ & $\bullet$ & $\bullet$ & $\bullet$ &  \\ \hline
Supply chain tracking & $\bullet$ & $\bullet$ & $\bullet$ & $\bullet$ & $\bullet$ &  \\ \hline
Inspections & $\bullet$ & $\bullet$ & $\bullet$ & $\bullet$ & $\circ$ &  \\ \hline
Storage and compensation programs & $\circ$ & $\circ$ & $\circ$ & $\circ$ & $\circ$ &  \\ \hline
Satellite and aerial surveillance & $\bullet$ & $\bullet$ & $\circ$ &  &  &  \\ \hline
Electrical Power Monitoring & $\bullet$ & $\bullet$ &  &  &  &  \\ \hline
Human intelligence and interviews & $\circ$ & $\circ$ & $\circ$ & $\circ$ & $\circ$ &  \\ \hline
Firmware-based chip tracking &  &  &  & $\circ$ & $\circ$ &  \\ \hline
Sting operations &  &  &  &  &  &  \\ \hline
Black market monitoring &  & $\bullet$ & $\bullet$ & $\bullet$ & $\bullet$ &  \\ \hline
Whistleblower programs & $\bullet$ & $\bullet$ & $\bullet$ & $\bullet$ & $\bullet$ &  \\ \hline
\end{tabular}%
}
\end{table}
\subsubsection{Staged Implementation of Chip Consolidation}
\begin{figure}
    \centering
    \includegraphics[width=0.9\linewidth]{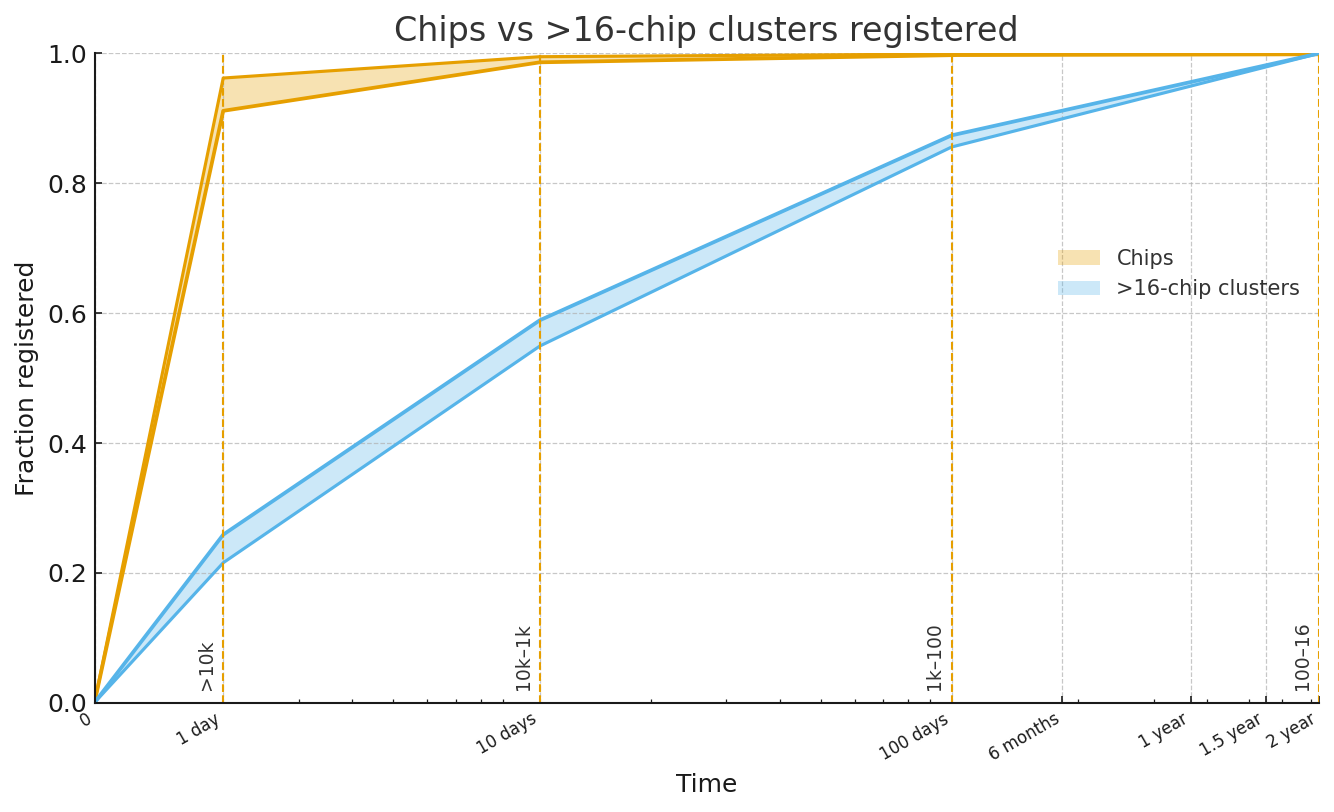}
    \caption{The timelines for locating clusters. Although it takes longer to register the majority of clusters, the majority of chips are quickly registered. This figure assumes that the size of AI clusters follows a Pareto distribution fit to data from \citet{pilz2025trends}.}
    \label{fig:chip_cluster_round_up}
\end{figure}

The consolidation of AI chips will proceed in phases, prioritizing the largest clusters first. Large clusters are the most urgent to secure, as they can most quickly reach training thresholds. They are also the easiest to detect due to their physical footprint, power consumption, extensive personnel requirements, and reliance on standard chip supply chains. Due to the current distribution of AI chips and AI supercomputers \citep{pilz2025trends}, the vast majority of AI chips are housed in relatively few facilities. By focusing on the largest clusters first, it will be fast to secure the majority of chips, even if it will take longer to register all covered clusters (Figure~\ref{fig:chip_cluster_round_up}).

\subparagraph{Pre-Agreement Preparation}
Before the international agreement enters into force, states will have begun tracking domestic compute infrastructure and preparing for rapid consolidation. Domestically, authorities should already be monitoring the location of clusters above 1,000 H100-equivalents, with particular focus on facilities exceeding 10,000 units.

States may engage in voluntary disclosure of major compute facilities to build mutual confidence. Intelligence gathering through satellite imagery, black market monitoring, and other means will also provide verification during this period.

Practical preparation includes establishing facilities to receive relocated chips, installing required security infrastructure (tamper-proof cameras, bandwidth limiters), and training personnel for the consolidation process. AI companies aware of the impending agreement can preemptively implement security measures, allowing inspectors to verify rather than install monitoring systems. These efforts can help ensure minimal disruption to everyday use of AI systems.

\subparagraph{Day 1: >10,000 H100-equivalent Clusters}
The immediate priority is identifying and beginning monitoring of all clusters exceeding 10,000 H100-equivalents. These facilities can typically continue operating in place with added monitoring, avoiding disruption to inference operations. While some may eventually require relocation away from population centers, this can happen gradually rather than all at once.

All entities must report clusters above the CCC threshold to domestic authorities. Personnel from the domestic authorities and the international verification effort deploy to major data centers to prevent any chips being smuggled out, supported by aerial surveillance to detect suspicious activity. A firmware update requirement takes effect, mandating location tracking for any chips receiving performance or security patches.

\subparagraph{Day 10: >1,000 H100-equivalent Clusters}
By this point, authorities should have located and begun monitoring all clusters above 1,000 H100-equivalents. Domestic efforts rely heavily on supply chain tracking, including engagement with enterprise server vendors, OEM companies, and their sales networks. Interviews with data center construction and maintenance personnel provide additional verification. Facilities suspected of housing significant compute undergo physical inspection.

International verification efforts intensify, with continual monitoring of 10,000+ unit clusters, verification of supply chain data through selective interviews, and investigation of all sites with especially high or anomalous power consumption (including aluminum smelters, desalination plants, and other high-draw facilities).

\subparagraph{Day 100: >100 H100-equivalent Clusters}
All clusters exceeding 100 H100-equivalents should now be under monitoring. Physical consolidation continues, with clusters below 1,000 units being physically relocated to secured facilities. International inspectors accompany all major chip transfers to ensure accurate accounting.

By this milestone, all covered clusters should be reported to domestic authorities and this information shared with international verification efforts. Supply chain investigations expand to include AI startups and research organizations that might possess significant compute.

\subparagraph{Year 2: All Covered Chip Clusters}
Over the next two years, the domestic consolidation effort shifts focus to tracking clusters between 16 and 100 H100-equivalents, ensuring comprehensive coverage of all CCCs.

International verification transitions to sustained intelligence gathering, primarily aimed at detecting potential state-run secret projects. Ongoing verification relies on continuous satellite monitoring, regular facility inspections, and interviews with industry personnel. This sustained effort ensures that the initial consolidation remains intact and that no significant number of existing chips fall outside of the monitoring framework.

\end{document}